\documentclass[10pt,journal,twoside]{IEEEtran}
\usepackage{amssymb}
\usepackage{amsthm}
\usepackage{multirow}
\usepackage{inputenc}
\usepackage{enumerate}
\usepackage{enumitem}
\usepackage{textcomp}
\usepackage{listings}
\usepackage{array}
\usepackage[switch,mathlines]{lineno}
\usepackage{soul}
\usepackage{xfrac}
\usepackage{graphicx}
\usepackage{caption}
\usepackage{cite}
\usepackage[cmex10]{amsmath}
\usepackage{xcolor}
\usepackage{colortbl}
\usepackage{cancel}
\usepackage{siunitx}
\usepackage{nomencl}
\usepackage{etoolbox} % 添加此包以支持 \ifstrequal 命令
\usepackage{multirow}
\usepackage{svg}
\usepackage{doi}
\usepackage{hyperref}
\usepackage{amsmath}
\usepackage{subfig}
\usepackage{verbatim}
\usepackage{etoolbox}
\usepackage[ruled,norelsize,vlined,linesnumbered]{algorithm2e}
\SetKwProg{Fn}{Function}{:}{end} % 定义函数环境
\makeatletter
\newcommand{\removelatexerror}{\let\@latex@error\@gobble}
\makeatother

\makeatletter
\def\footnoterule{\kern-3\p@
  \hrule \@width 3.3in \kern 2.6\p@} % the \hrule is .4pt high
\makeatother

\makenomenclature

\usepackage{algpseudocode}

\usepackage{stackengine}

\usepackage{algpseudocode}
\usepackage{CJK}

\usepackage{hyperref}
\makeatletter
\def\UrlAlphabet{%
      \do\a\do\b\do\c\do\d\do\e\do\f\do\g\do\h\do\i\do\j%
      \do\k\do\l\do\m\do\n\do\o\do\p\do\q\do\r\do\s\do\t%
      \do\u\do\v\do\w\do\x\do\y\do\z\do\A\do\B\do\C\do\D%
      \do\E\do\F\do\G\do\H\do\I\do\J\do\K\do\L\do\M\do\N%
      \do\O\do\P\do\Q\do\R\do\S\do\T\do\U\do\V\do\W\do\X%
      \do\Y\do\Z}
\def\UrlDigits{\do\1\do\2\do\3\do\4\do\5\do\6\do\7\do\8\do\9\do\0}
\g@addto@macro{\UrlBreaks}{\UrlOrds}
\g@addto@macro{\UrlBreaks}{\UrlAlphabet}
\g@addto@macro{\UrlBreaks}{\UrlDigits}
\makeatother
\usepackage[cmex10]{amsmath}
\usepackage{bm}
\usepackage{color}
\usepackage{amsmath,amsthm,amssymb,amsfonts}
\usepackage{flushend}
\usepackage{float}

\newtheorem{remark}{Remark}

\usepackage{arydshln} %to include dashed lines in arrays
\usepackage{hyperref}
\hypersetup{
     colorlinks   = true,
     linkcolor    = blue,
     citecolor    = blue,
     urlcolor     = blue
}
\makeatletter
\newcommand*{\transpose}{%
  {\mathpalette\@transpose{}}%
}
\newcommand*{\@transpose}[2]{%
  \raisebox{\depth}{$\m@th#1\intercal$}%
}
\makeatother

\usepackage{soul}
\usepackage{tikz}
\usepackage{booktabs}% http://ctan.org/pkg/booktabs
\usepackage{tabularx}% http://ctan.org/pkg/tabularx
\usepackage{setspace}

\newtheorem{definition}{Definition}

\DeclareMathOperator{\E}{\mathbb{E}}
\ifCLASSINFOpdf
\else
\fi
\renewcommand\nomgroup[1]
{
  \ifthenelse{\equal{#1}{A}}{\item[\textit{A. Abbreviations}]}{%
  \ifthenelse{\equal{#1}{B}}{\item[\textit{B. Sets}]}{
   \ifthenelse{\equal{#1}{C}}{\item[\textit{\textcolor{black}{C. Parameters}}]}{
  \ifthenelse{\equal{#1}{D}}{\item[\textit{\textcolor{black}{D. Variables}}]} \\
  }
  }
  }
  }
\begin{document}
\renewcommand{\ttdefault}{cmtt}
\bstctlcite{IEEEexample:BSTcontrol}
\SetKwComment{Comment}{/* }{ */}
\renewcommand\qedsymbol{$\blacksquare$}
% \graphicspath{{figure/}}

% ******************************************************************************
\title{\huge Integrating Building Thermal Flexibility Into Distribution System: A Privacy-Preserved Dispatch Approach}

\author{
{
Shuai Lu,~\IEEEmembership{Member, IEEE},
Zeyin Hou,~\IEEEmembership{Student Member, IEEE},
Wei Gu,~\IEEEmembership{Senior Member, IEEE}, \\
Yijun Xu,~\IEEEmembership{Senior Member, IEEE}
}

\thanks{The work was supported in part by the National Natural Science Foundation of China (52325703, 52207080), and in part by the Zhishan Young Scholar Support Program, Southeast University (2242024RCB0044). (\emph{Corresponding author: Wei Gu}).}

\thanks{S. Lu, Z. Hou, W. Gu, and Y. Xu are with the  Electrical Engineering Department, Southeast University, Nanjing, Jiangsu 210096, China, (e-mail: \href{mailto:shuai.lu.seu@outlook.com}{shuai.lu.seu@outlook.com}; \href{mailto:220222681@seu.edu.cn}{220222681@seu.edu.cn}; \href{mailto:wgu@seu.edu.cn}{wgu@seu.edu.cn}; \href{mailto:yijunxu@seu.edu.cn}{yijunxu@seu.edu.cn}).
}
}

\markboth{A\MakeLowercase{ccepted for publication in} IEEE Transactions on Industrial Informatics}
{LU, \MakeLowercase{et al.:} Integrating Building Thermal Flexibility Into Distribution System: A Privacy-Preserved Dispatch Approach}
\maketitle

\vspace{-1cm}
\begin{abstract}
% Necessity: 
The inherent thermal storage capacity of buildings brings considerable thermal flexibility to the heating/cooling loads, which are promising demand response resources for power systems.
It is widely believed that integrating the thermal flexibility of buildings into the distribution system can improve the operating economy and reliability of the system.

% Problem: 
However, the private information of the buildings needs to be transferred to the distribution system operator (DSO) to achieve a coordinated optimization, bringing serious privacy concerns to users.
% Method:
Given this issue, we propose a novel privacy-preserved optimal dispatch approach for the distribution system incorporating buildings. Using it, the DSO can exploit the thermal flexibility of buildings without accessing their private information, such as model parameters and indoor temperature profiles.
Specifically, we first develop an optimal dispatch model for the distribution system integrating buildings, which can be extended to other storage-like flexibility resources.
Second, we reveal that the privacy-preserved integration of buildings is a joint privacy preservation problem for both parameters and state variables and then design a privacy-preserved algorithm based on transformation-based encryption, constraint relaxation, and constraint extension techniques.
Besides, we implement a detailed privacy analysis for the proposed method, considering both semi-honest adversaries and external eavesdroppers.
% Effect: 
Case studies demonstrate the accuracy, privacy-preserved performance, and computational efficiency of the proposed method.
\end{abstract}

\IEEEpeerreviewmaketitle
\vspace{-0.2cm}

% abbreviations
\nomenclature[A]{ATDM}{Aggregate thermal dynamic model}
\nomenclature[A]{BLA}{Building load aggregator}
\nomenclature[A]{DSO}{Distribution System Operator}
\nomenclature[A]{CRT}{Constraint Relaxation Technique}
\nomenclature[A]{CET}{Constraint Extension Technique}
\nomenclature[A]{TE}{Transformation-based Encryption}
% set
\nomenclature[B]{$\mathbf{T}$}{Index set of time periods}
\nomenclature[B]{$\mathbf{M}$}{Index set of model order}
\nomenclature[B]{$\mathbf{J}$}{The index set of the branches}
\nomenclature[B]{$\mathbf{J}_k^+,\mathbf{J}_k^-$}{The index set of the upstream and downstream branches of the bus $k$}
\nomenclature[B]{$\mathbf{K}_j^+,\mathbf{K}_j^-$}{The index set of the initial and terminal buses of the branch $j$}
\nomenclature[B]{$\mathbf{K}$}{Index set of buses}

% parameter
\nomenclature[C]{$\alpha_k^m$, $\beta_k^m$, $\gamma_k^t$}{Parameters of the ATDM of BLA $k$}
\nomenclature[V]{$\bm{W}_k,\bm{V}_k$}{Invertible random matrices privately owned by the BLA on bus $k$}
\nomenclature[C]{$\bar{\tau}_k$, $\underline{\tau}_k$}{Upper (lower) temperature limits of BLA $k$}
\nomenclature[C]{\(M\)}{Model order}
\nomenclature[C]{\(T\)}{The total number of periods}
\nomenclature[V]{$\xi_k^i$}{The aggregate coefficient of the building zone $i$ in the BLA $k$}
\nomenclature[C]{$c_{grid,b}^t$, $c_{grid,s}^t$}{The purchase and sale price of the electricity}
\nomenclature[C]{$c_{bt}^k$, $c_{res}^k$}{The operation and maintenance cost of the battery and renewable power generation}
\nomenclature[C]{$\bar{P}_{tie}^t$}{The power limit of the tie line between the distribution system and the main grid}
\nomenclature[C]{$\bar{P}_{res}^k$, $\bar{Q}_{res}^k$}{Active and reactive power limits of the renewable energy generation on bus $k$}
\nomenclature[C]{$\bar{P}_{bt,chr}^{k}$, $\bar{P}_{bt,dis}^{k}$}{The charging and discharging power limit of the battery on bus $k$}
\nomenclature[C]{$\eta_{chr}^k$, $\eta_{dis}^k$}{The charging and discharging efficiency of the battery on bus $k$}
\nomenclature[C]{$\bar{E}_{bt}^k$, $\underline{E}_{bt}^k$}{The upper and lower limits of the energy storage level on bus $k$}
\nomenclature[C]{$r_{br}^j$, $x_{br}^j$}{The resistance and reactance of the branch $j$}
\nomenclature[C]{$V_0$}{Reference voltage magnitude}
\nomenclature[C]{$\bar{V}^k$, $\underline{V}^k$}{The upper and lower limits of the voltage magnitude on bus $k$}
\nomenclature[C]{$\bar{P}_{br}^j$}{The upper limit of active power on branch $j$}
\nomenclature[D]{$P_{load}^{k,t}$, $Q_{load}^{k,t}$}{The active and reactive loads on bus $k$}
%\nomenclature[V]{$\bm{\Lambda}_k$, $\bm{R}_k$, $\bm{S}_k$, $\bm{d}_k$, $\bm{D}_k$, $\bm_{E}_k$, $\bm{x}_k^{bd}$, $\bm{F}_k$, $\bm{G}_k$, $\bm{H}_k$, $\bm{e}_k$}{The parameter matrices of the BLA on bus $k$}
\nomenclature[C]{$\bar{x}_k$, $\underline{x}_k$}{The upper and lower limits of $\bm{x}_k$}
\nomenclature[C]{$\bar{Q}_{bt}^{k}$}{The reactive power upper limit of the battery on bus $k$}

% variable
\nomenclature[D]{$\tau_{in}^{k,i,t}$}{The indoor temperature of the building zone $i$ in the BLA $k$}
\nomenclature[D]{$\tilde{\tau}_k^t$}{The aggregate temperature of BLA $k$}
\nomenclature[D]{$P_{BLA}^{k,t}$}{The heating/cooling power of the BLA $k$}
\nomenclature[D]{$C_{grid}$}{Transaction cost with the grid}
\nomenclature[D]{$C_{om}$}{Operation and maintenance cost}
\nomenclature[D]{$P_{grid,b}^t$, $P_{grid,s}^t$}{Electrical power purchased from and sold to the grid}
\nomenclature[D]{$P_{bt,chr}^{k,t}$, $P_{bt,dis}^{k,t}$}{Charging and discharging power of the battery on bus $k$}
\nomenclature[D]{$P_{res}^{k,t}$, $Q_{res}^{k,t}$}{Active and reactive power of renewable power generation on bus $k$}
\nomenclature[D]{$\varepsilon_{grid,b}^t$, $\varepsilon_{grid,s}^t$}{Electricity purchase mode}
\nomenclature[D]{$Q_{bt}^{k,t}$}{Reactive power of the battery on bus $k$}
\nomenclature[D]{$E_{bt}^{k,t}$}{Energy storage level on bus $k$}
\nomenclature[D]{$\varepsilon_{bt,chr}^{k,t}$, $\varepsilon_{bt,dis}^{k,t}$}{Charging and discharging state of the battery on bus $k$}
\nomenclature[D]{$P_{inj}^{k,t}$, $Q_{inj}^{k,t}$}{Active (reactive) power injection on bus $k$}
\nomenclature[D]{$P_{br}^{j,t}$, $Q_{br}^{j,t}$}{The active and reactive power on branch $j$}
\nomenclature[D]{$V^{k,t}$}{The voltage magnitude on bus $k$}
\nomenclature[D]{$k,\ j$}{The index of the bus/branch}
%\nomenclature[V]{$\bm{x}_k$, $\bm{u}_k$}{The state vector and the vector of the BLA on bus $k$}
%\nomenclature[V]{$\tilde{\bm{x}}_k$}{The masked pseudo-state of $\bm{x}_k$}
%\nomenclature[V]{$\bm{\omega}_k$}{The slack variables of the BLA on bus $k$}
\printnomenclature[2.2cm]

\vspace{-0.2cm}
\section{Introduction}
\IEEEPARstart{A}{s} {THE} penetration of renewable power generation increases, flexible resources become increasingly important for the power system to balance electricity supply and demand, improve grid stability, save operating costs \textcolor{black}{and facilitate the low-carbon operation of power systems} \cite{wang2024robust}, \cite{zhong2023optimal}. 
Demand-side flexibility resources, including the heating/cooling loads of buildings, electric vehicles, and energy storage units, have shown great potential to promote the local assumption of renewable energy due to their superiority in adjustable capacity and response speed. As a representative flexible resource on the demand side, the heating/cooling loads of buildings, which account for a large percentage of the total electrical load, have considerable flexibility due to the inherent thermal storage capacity of buildings, i.e., thermal flexibility. Hence, they have received widespread attention from both academia and industry \cite{taha2017buildings}, \cite{hou2025robust}.

% 接下来这一段，应该介绍相关的研究
Much research has focused on integrating the thermal flexibility of buildings into power systems. 
Fontenot \emph{et al.} \cite{fontenot2019modeling} reviews the modeling and control of the building-integrated power system. \textcolor{black}{Guo \emph{et al.} \cite{guo2024optimal} proposes the data-driven energy hubs and thermal dynamics of pipeline networks to facilitate the utilization of thermal flexible resources.} Niu \emph{et al.} \cite{niu2019flexible} investigates the flexibility
potential of thermal storage in the energy system of a building. In summary, integrating buildings into the power system can improve the system's flexibility. However, the large number of buildings and their spatially dispersed distribution bring prohibitive computing and communication burdens, making it difficult for power systems to interact with them peer-to-peer directly.
Considering this, the building load aggregator (BLA) based operation framework is put forth, in which the BLA aggregates the heating/cooling loads of buildings to interact with power systems, acting as an interface between the power system operator and building users \cite{song2018state}, \cite{correa2019optimal}. Lu \emph{et al.} \cite{lu2021data} proposes the aggregate thermal dynamic model to offer an equivalent and computationally efficient building model, serving as an interface to integrate into the power system. Zhang \emph{et al.} \cite{zhang2019day} propose a day-ahead scheduling framework for integrated electricity and district heating systems considering the aggregate model of buildings. Tang \emph{et al.} \cite{tang2022multi} propose an optimal coordinated day-ahead bidding strategy for the BLA to maximize the economic benefits in the electricity market. This article also adopts the BLA to characterize the thermal flexibility of buildings. \textcolor{black}{Given that most contemporary power systems rely on centralized computing paradigms for their simplicity, this paper focuses on the centralized dispatch problem.}
 
However, integrating the thermal flexibility of buildings into power systems can bring serious privacy concerns \cite{yang2021survey} to the BLA and occupants. This is because the DSO must acquire private information about buildings to implement a coordinated optimization. Specifically, the high-fidelity model parameters of the buildings's aggregate model, the business secrets of BLA, are disclosed to the DSO. Moreover, during optimization, the DSO can infer the (aggregate) indoor temperature profiles, which may contain occupants' living patterns. 
Therefore, it is urgent to propose a privacy-preserved computation method for integrating the thermal flexibility of buildings into power systems for coordinated scheduling. This is chosen as the focus of our article. 

Here, numerous methods have been investigated for the privacy-preserved computation problem. 
Among them, the ADMM is a representative distributed algorithm that partially protects privacy because agents exchange the intermediate information like the gradients instead of the original sensitive information \cite{zhang2024admm}. However, attackers may infer private information by combining the exchanged information during each iteration, raising serious privacy concerns \cite{zhang2018admm}. 
The differential privacy (DP) is a commonly used metric to assess the privacy preservation effect realized by introducing carefully calibrated noise \cite{shang2021differentially}. \textcolor{black}{Liu \emph{et al.} \cite{liu2024differentially} propose 
 a privacy-preserved distributed Nash-seeking strategy based on differential privacy to safeguard against potential attacks.} Hunag \emph{et al.} \cite{huang2019dp} propose a novel differentially private ADMM-based distributed learning algorithm named DP-ADMM. Dvorkin \emph{et al.} \cite{dvorkin2020differentially} initially introduced the privacy-preserving optimal power flow mechanism for the distribution grids based on the DP framework. Nevertheless, the DP method introduces noise, damaging the computation accuracy. 
Alternatively, homomorphic encryption (HE) enables computations to be implemented on the encrypted data through cryptographic means\cite{wu2021privacy}, \cite{shoukry2016privacy}.

Despite the privacy-preservation performance, the prohibitive computation load restricts the practical application of the HE.
Besides, transformation-based encryption (TE) is a lightweight and computationally precise method that masks the true data via equivalent linear mapping.

Xin \emph{et al.}  \cite{xin2017information} propose the transformation-based encryption for the cloud-based energy management system. Jia \emph{et al.} \cite{jia2022chance} use the TE method to solve the chance-constrained optimal power flow without disclosing the confidential data of agents. However, this method requires an independent third party like the cloud or independent system operator to implement computing \cite{xin2017information}, \cite{wu2017transformation}, \cite{gonccalves2021critical}.

% Research gaps & Difficulties
Despite the advances made in the privacy-preservation calculation, establishing a privacy-preserved method to integrate the thermal flexibility of buildings into power systems receives little attention. Besides, the existing privacy-preserved methods are not directly applicable to this problem. Given the computational and communication challenges of the distributed approach, we consider optimal dispatch as a centralized optimization problem in this article, with the DSO as the decision-maker. This fails the privacy-preservation methods based on the distributed manner like the ADMM. Moreover, the inaccuracy and heavy computation burden of the DP and HE limit their practicality in this problem. Besides, the integration essentially leads to variable coupling between the decision-maker, i.e., the DSO, and the private information providers, i.e., the BLA. This makes the decision-maker, not an independent third party, invalidating the traditional TE method.

 Therefore, we are motivated to develop a privacy-preserved optimal dispatch approach for the distribution system with the integration of thermal flexibility of buildings. Based on this method, the DSO can implement the optimal dispatch without accessing the sensitive information of buildings, including the model parameters and (aggregate) indoor temperature profiles.
 
 The main contributions are summarized as follows.
 
\textcolor{black}{(1)  We propose a privacy-preserved centralized computing algorithm for the optimal dispatch problem of the distribution system integrating BLAs. This method can solve the optimal dispatch problem without exposing the privacy information of BLAs, i.e., the model parameters and state variables.}

\textcolor{black}{(2) We conduct a detailed privacy analysis for the proposed privacy-preserved method. We reveal that the essence of privacy inference is solving multivariate equations. This ensures the proposed method can protect the private information of one individual BLA from potential adversaries, including semi-honest adversaries like DSO and other BLAs as well as external eavesdroppers.}

\textcolor{black}{The rest of this paper is organized as follows. Section \ref{Section II} introduces the optimal dispatch problem considering BLAs. Section \ref{Section III} proposes the privacy-preserved algorithm for the dispatch model. Section \ref{Section IV} gives the numerical simulation results, and Section \ref{Section V} concludes this paper and prospects for future research.}

%The remainder of this paper is organized as follows. Section \ref{Section II} introduces the optimal dispatch dispatch problem of the distribution system with the integration of buildings. Section \ref{Section III} proposes the privacy-preserved algorithm for the coordinated optimization and carries out the detailed privacy analysis. Section \ref{Section IV} gives the numerical simulation results, and Section \ref{Section V} concludes the paper.

\vspace{-0.1cm}
\section{Problem Description}
\label{Section II}
\begin{figure}[t]
    \centering
    \includegraphics[width=0.9\columnwidth]{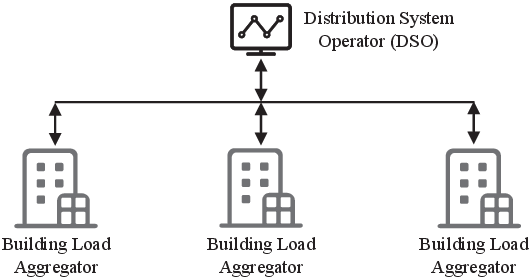}
    \setlength{\abovecaptionskip}{-5pt}
    \vspace{0.4cm}
    \caption{\textcolor{black}{The diagram for the dispatch model.}}
    \label{dispatch framework}
\end{figure}

In this section, we illustrate the optimal dispatch problem of the distribution system with the integration of building thermal flexibility.
\textcolor{black}{The dispatch framework is presented in Fig. \ref{dispatch framework}, where the DSO acquires the model data from BLAs, optimizes the economic dispatch schedule based on this information, and subsequently issues control dispatch results to BLAs.}

\vspace{-0.3cm}
\subsection{Optimal Dispatch Model}
\label{foundamental optimal dispatch problem}
%The optimal dispatch of the distribution system aims to obtain the optimal outputs of devices and power flow in the network so that the system can operate most economically. The BLA is an effective interface for the heating/cooling loads of buildings to interact with the power system. 

\subsubsection{The BLA Model}
In this paper, we use the aggregate indoor thermal dynamic model (ATDM) proposed in \cite{lu2021data} to describe the BLA. In the ATDM, the aggregate temperature is selected as the characteristic state variable to represent the global state of a building cluster, defined as
\begin{subequations}
\begin{equation}
    \label{the definition of the aggregate temperature}
    \tilde \tau_{in}^{k,t} =  \sum \limits_{i \in {\mathbf{I}_k}} \xi_k^i\tau _{in}^{k,i,t}, \;
    \forall k \in {{\mathbf{K}}},
\end{equation}
wherein $\mathbf{I}_k$ is the building zone set in the building cluster $k$, the aggregate coefficient $\xi_k^i$ denotes the contribution of each building zone to the global state, which should satisfy
\begin{equation}
    \label{the constraints for aggregation coefficients}
    \mathop \sum \limits_{i \in {\mathbf{I}_k}} \xi_k^i = 1,\;
    \xi_k^i \geq 0,\; 
    \forall i \in {{\mathbf{I}_k}},\; 
    \forall k \in \mathbf{K}.
\end{equation}

Based on this, the ATDM describes the aggregate state equation using a linear model as follows
\begin{equation}
\begin{aligned}  % 用aligned而不是gathered
    \label{BLA model1-1}
     \tilde \tau_{in}^{k,t} = \underset{m\in\mathbf{M}\backslash\{0\}} {\sum} \alpha_k^m  \tilde \tau_{in}^{k,t-m} + 
     \underset{m\in\mathbf{M}}{\sum} \beta_k^m P_{BLA}^{k,t-m} 
     + \gamma_k^t, \\ 
     \forall k \in \mathbf{K},\forall t \in {\mathbf{T}},
\end{aligned}
\end{equation}
wherein the parameters $\alpha_k^m$, $\beta_k^m$, and $\gamma_k^t$ can be obtained by the BLA through parameter estimation algorithm, e.g., the least square estimation \cite{lu2021data}.
Besides, the temperature constraints are needed to ensure the thermal comfort of users, as follows
\begin{equation}
    \label{BLA model1-2}
    \underline {\tau}_k\leq \tilde \tau_{in}^{k,t} \leq \bar{\tau}_k.
\end{equation}
\end{subequations}

Finally, the BLA model consists of the ATDM equation \eqref{BLA model1-1} and the thermal comfort constraint \eqref{BLA model1-2}. 
\subsubsection{The Optimal Dispatch Model}
We choose the operational cost as the objective function of the optimal dispatch model of the distribution system, as follows 
\begin{subequations}
\label{objective of ED}
\begin{equation}
    \label{total cost of the DSO}
    \min \ C_{grid} + C_{om},
\end{equation}
wherein the electricity transaction cost $C_{grid}$ and the operational maintenance cost $C_{om}$ are separately defined as 
\begin{equation}
    \label{cost of purchasing electricity from the grid}
    C_{grid} = \sum \limits_{t \in {\mathbf{T}}} \left( c_{grid,b}^t P_{grid,b}^t-c_{grid,s}^t P_{grid,s}^t \right) \Delta t,
\end{equation}
\begin{equation}
    \label{cost of operating and matennance}
    C_{om} =\sum \limits_{t \in {\mathbf{T}}}  \sum \limits_{k\in \mathbf{K}} \left( c_{bt}^k(P_{bt,chr}^{k,t}+P_{bt,dis}^{k,t})+c_{res}^k P_{res}^{k,t} \right) \Delta t.
\end{equation}
\end{subequations}

The operational constraints include the BLA model (i.e., \eqref{BLA model1-1}-\eqref{BLA model1-2}) and the constraints of the grid, as follows 
\begin{subequations}
    \label{constraints-ED}
\begin{equation}
    \label{constraint-grid exchange-buy}
    0\leq P_{grid,b}^t \leq \varepsilon_{grid,b}^t  \bar{P}_{tie}^{t}, \; t \in {\mathbf{T}},
\end{equation}
\begin{equation}
    \label{constraint-grid exchange-sell}
    0\leq P_{grid,s}^t \leq \varepsilon_{grid,s}^t  \bar{P}_{tie}^{t}, \; t \in {\mathbf{T}},
\end{equation}
\begin{equation}
    \label{constraint-grid exchange-state exclusion}
    \begin{aligned}
    \varepsilon_{grid,b}^{t}+\varepsilon_{grid,s}^{t} \leq 1,\;
      t \in {\mathbf{T}},
    \end{aligned}
\end{equation}
\begin{equation}
    \label{constraint-grid exchange-integer state}
    \begin{aligned}
\varepsilon_{grid,b}^{t},\ \varepsilon_{grid,s}^{t}\in \{0,1\}, \;
      t \in {\mathbf{T}},
    \end{aligned}
\end{equation}
\begin{equation}
\label{constraint-RES-P-UpperLimit}
    0 \leq P_{res}^{k,t} \leq 
    \bar{P}_{res}^{k,t},\ k \in {{\mathbf{K}}},\;t \in {\mathbf{T}},
\end{equation}
\begin{equation}
    \label{constraint-RES-Q-UpperLimit}
    0\leq Q_{res}^{k,t}\leq \bar{Q}_{res}^{k,t},\ k \in {{\mathbf{K}}},\;t \in {\mathbf{T}},
\end{equation}
\begin{equation}
    \label{constraint-battery-ChargePower-UpperLimit}
    0\leq P_{bt,chr}^{k,t} \leq \varepsilon_{bt,chr}^{k,t} \bar{P}_{bt,chr}^{k,t}, k \in {\mathbf{K}},\;t \in {\mathbf{T}},
\end{equation}
\begin{equation}
    \label{constraint-battery-DischargePower-UpperLimit}
    0\leq P_{bt,dis}^{k,t}\leq \varepsilon_{bt,dis}^{k,t} \bar{P}_{bt,dis}^{k,t},\ k\in {\mathbf{K}},\;t \in {\mathbf{T}},
\end{equation}
\begin{equation}
    \label{constraint-battery-ReactivePower-Limit}
    -\bar{Q}_{bt}^{k}\leq Q_{bt}^{k,t}\leq \bar{Q}_{bt}^{k}, k \in {\mathbf{K}},\;t \in {\mathbf{T}},
\end{equation}
\begin{equation}
    \begin{aligned}
        \label{constraint-battery-CapacityChange}
    E_{bt}^{k,t} = (1-\sigma_{bt}^{k})E_{bt}^{k,t-1}+\eta_{chr}^k P_{bt,chr}^{k,t}-P_{bt,dis}^{k,t}/\eta_{dis}^{k,t},   \\
     k \in {\mathbf{K}}, \;
      t \in {\mathbf{T}},
      \end{aligned}
\end{equation}
\begin{equation}
\label{constraint-battery-CapacityLimit}
    \underline{E}_{bt}^{k}\leq E_{bt}^{k,t}\leq  \bar{E}_{bt}^{k}, \;
    k \in {\mathbf{K}}, \;
      t \in {\mathbf{T}},
\end{equation}
\begin{equation}
    \label{constraint-battery-IntegerLimit}
    \begin{aligned}
    \varepsilon_{bt,chr}^{k,t}+\varepsilon_{bt,dis}^{k,t} \leq 1,\ \varepsilon_{bt,chr}^{k,t},\varepsilon_{bt,dis}^{k,t}\in \{0,1\}, \\
    k \in {\mathbf{K}}, \;
      t \in {\mathbf{T}},
    \end{aligned}
\end{equation}
\begin{equation}
    \label{constraint-bus-ActivePower}
    P_{inj}^{k,t} + \sum \limits_{j\in \mathbf{J}_k^+} P_{br}^{j,t}=\sum \limits_{j\in \mathbf{J}_k^-} P_{br}^{j,t}, \;
     k \in {\mathbf{K}}, \;
      t \in {\mathbf{T}}, \;
\end{equation}
\begin{equation}
    \label{constraint-bus-ReactivePower}
    Q_{inj}^{k,t} + \sum \limits_{j\in \mathbf{J}_k^+} Q_{br}^{j,t}=\sum \limits_{j\in \mathbf{J}_k^-}  Q_{br}^{j,t}, \; 
     k \in {\mathbf{K}}, \;
      t \in {\mathbf{T}},
\end{equation}
\begin{equation}
\label{constraint-VoltageDrop}
\begin{aligned}
     V^{l,t}=V^{k,t}-(r_{br}^{j}P_{br}^{j,t}+x_{br}^{j}Q_{br}^{j,t})/V_0,\;
j\in \mathbf{J}, \\
k \in \mathbf{K}_j^{+}, \;
l \in \mathbf{K}_j^{-}, 
t \in {\mathbf{T}},
\end{aligned}
\end{equation}
\begin{equation}
    \label{constraint-VoltageRange}
    \underline{V}^k \leq V^{k,t}\leq  \bar{V}^k, \;
    k \in {\mathbf{K}}, \;
      t \in {\mathbf{T}},
\end{equation}
\begin{equation}
    \label{constraint-BranchActivePowerRange}
    -\bar{P}_{br}^{j}\leq P_{br}^{j,t}\leq \bar{P}_{br}^{j}, \;
    j \in {\mathbf{J}}, \;
      t \in {\mathbf{T}},
\end{equation}
\begin{equation}
\begin{aligned}
    \label{constraint-ActivePowerInjection}
    P_{inj}^{k,t}=P_{res}^{k,t}+P_{bt,dis}^{k,t}-P_{bt,chr}^{k,t}-P_{load}^{k,t} - P_{BLA}^{k,t}, \\
    k \in {\mathbf{K}}, \;
      t \in {\mathbf{T}},
      \end{aligned}
\end{equation}
\begin{equation}
\begin{aligned}
    \label{constraint-ReactivePowerInjection}
    Q_{inj}^{k,t}= Q_{res}^{k,t}+Q_{bt,dis}^{k,t}-Q_{bt,chr}^{k,t} - Q_{load}^{k,t}, \\
    k \in {\mathbf{K}}, \;
      t \in {\mathbf{T}}.
\end{aligned}
\end{equation}
\end{subequations}

The constraints in \eqref{constraints-ED} are related to the distribution system, wherein \eqref{constraint-grid exchange-buy}-\eqref{constraint-grid exchange-integer state} are the constraints of the point of the common coupling between the main grid and the distribution system, \eqref{constraint-RES-P-UpperLimit}-\eqref{constraint-RES-Q-UpperLimit} are the constraints of renewable power generation, \eqref{constraint-battery-ChargePower-UpperLimit}-\eqref{constraint-battery-IntegerLimit} are constraints of the energy storage batteries, \eqref{constraint-bus-ActivePower}-\eqref{constraint-bus-ReactivePower} are the constraints of the bus active/reactive power balance, \eqref{constraint-VoltageDrop} is the linearized Distflow branch power model \cite{rigo2022iterative}, \cite{zhang2016robust}, \eqref{constraint-VoltageRange}-\eqref{constraint-BranchActivePowerRange} are the constraints of voltage amplitude and branch power respectively, and \eqref{constraint-ActivePowerInjection}-\eqref{constraint-ReactivePowerInjection} are the expression of active/reactive injection power, depicting the coupling relationship between the distribution system and the BLAs. 
\textcolor{black}{It is worth mentioning that the uncertainty of renewable energy generation significantly impacts economic dispatch. Methods including robust optimization or stochastic optimization could be employed to address such optimal dispatch problems considering uncertainty.}

\subsection{Compact Form}
In the following, we transform the above model into a compact form for the convenience of subsequent analysis.
\subsubsection{Compact Form of the BLA Model}
First, We define the matrix sequences $\bm{\Lambda}^m \in\mathbb{R}^{T\times T}$ ($m=0,1\cdots,M$), wherein the element in the $i$-th row and $j$-th column is defined as
$$
    \label{definition-lambda matrix }
    \left(\Lambda^m\right)_{i,j} \triangleq
    \begin{cases}
        1 & i-j=m, \\
        0 & i-j\neq m.
    \end{cases}
$$

As a special case, $\bm{\Lambda}^0$ is essentially the $T$-dimensional identity matrix $\mathbf{I}_T$.
Then, we use the state variable $x_k^t$ to represent the aggregate temperature variable, i.e., $\tilde \tau_{in}^{k,t}$, the control variable $u_k^t$ to represent the power injected into the BLA $k$, i.e., $P_{BLA}^{k,t}$, and $\bar{x}_k$ (or $\underline{x}_k$) to represent the upper (or lower) limit of the aggregate temperature $\bar{\tau}_k$ (or $ \underline{\tau}_k$).
 
 Then, we stack the state variables and the control variables to form the state vector $\bm{x}_k$ and control vector $\bm{u}_k$, as 
$$
\bm{x}_k \triangleq [x_k^1,x_k^2,\cdots,x_k^T]^ \top \in \mathbb{R}^{T\times 1}.
$$
$$
\bm{u}_k \triangleq [u_k^1,u_k^2,\cdots,u_k^T]^\top \in \mathbb{R}^{T\times 1},
$$
Then, the equation \eqref{BLA model1-1} can be equivalently transformed as 
\begin{subequations}
\begin{equation}
\bm{x}_k = \underset{m\in\mathbf{M} \backslash \{0\}}{\sum}\alpha_k^m\bm{\Lambda}^m \bm{x}_k+\underset{m\in\mathbf{M}}{\sum} \beta_k^m \bm{\Lambda}^m \bm{u}_k + \bm{d}_k,
\end{equation}
wherein $\bm{d}_k \in \mathbb{R}^T$ is a constant vector depending on the paramters, the historical state variables, and the historical control variables. Its $t$-th element is
\begin{equation*}
    \label{general state equation-residue-matrix form}
    d_k^t = 
    \begin{cases}
        \underset{t\leq m\leq M}{\sum}(\alpha_k^m x_k^{t-m}+\beta_k^m u_k^{t-m})+\gamma_k^t, \;
        1\leq t\leq M, \\
        \quad \gamma_k^t, \;
        \ M+1\leq t\leq T.
    \end{cases}
\end{equation*}

Also, we define the coefficient matrices $\bm{R}_k$ and $\bm{S}_k$ as 
$$ \bm{R}_k = \mathbf{I}_T - \underset{m\in\mathbf{M}\backslash\{0\}}{\sum}\alpha_k^m\bm{\Lambda}^m\in\mathbb{R}^{T\times T}, $$
$$ \bm{S}_k = \underset{m\in\mathbf{M}}{\sum}\beta_k^m\bm{\Lambda}^m\in\mathbb{R}^{T\times T}, $$

Then, the BLA model \eqref{BLA model1-1}-\eqref{BLA model1-2} can be reformulated into the matrix form, as
\begin{equation}
    \label{BLA model2-1-original}
    \bm{R}_k \bm{x}_k + \bm{S}_k \bm{u}_k=\bm{d}_k,
\end{equation}
\begin{equation}
    \label{BLA model2-2-original}
    \underline{x}_k\cdot \bm{1}_T \leq \bm{x}_k\leq\bar{x}_k\cdot \bm{1}_T.
\end{equation}
\end{subequations}

\subsubsection{Compact Form of the Optimal Dispatch Model}
We denote the variables in \eqref{constraints-ED}, except for $P_{BLA}^{k,t}$, as the distribution system variables $\bm{z}$, and denote the feasible region formed by the constraints \eqref{constraint-grid exchange-buy}-\eqref{constraint-BranchActivePowerRange} and \eqref{constraint-ReactivePowerInjection} as $\mathbf{Z}$. Then, the constraints of the distribution system are  
\begin{equation*}
    \bm{z}\in \mathbf{Z}  \triangleq \{\eqref{constraint-grid exchange-buy}-\eqref{constraint-BranchActivePowerRange}, \eqref{constraint-ReactivePowerInjection} \}.
\end{equation*}

Define $\bm{u} \triangleq  \left[ \bm{u}_1^\top,\cdots,\bm{u}_{K}^\top \right ]^\top\in\mathbb{R}^{TK\times 1}$, and the power coupling constraints between the distribution system and the BLA can be unified as 
\begin{equation*}
    \bm{A}\bm{z}+\bm{u} = \bm{0},
\end{equation*}
wherein the matrix $\bm{A}$ denotes the connecting relationship between the buses and BLAs, which the DSO knows.
Also, the feasible region of the control vector $\bm{u}_k$ , denoted as $\mathbf{U}_k$, can be represented by
\begin{equation*}
\begin{split}
\label{the feasible random of U_k-1}
\mathbf{U}_k = \{ & \bm{u}_k|\ \bm{R}_k\bm{x}_k +  \bm{S}_k\bm{u}_k=\bm{d}_k, \\
        & \underline{x}_k\cdot \bm{1}_T \leq \bm{x}_k\leq\bar{x}_k\cdot \bm{1}_T\}.
\end{split}
\end{equation*}

Define $\bm{x} \triangleq  \left[ \bm{x}_1^\top,\cdots,\bm{x}_{K}^\top \right ]^\top\in\mathbb{R}^{TK\times 1}$. Finally, the compact form of the dispatch model is as  
\begin{equation}
    \label{P0}
    \begin{split}
        \ \underset{\bm{z},\bm{x},\bm{u}}{\min} & \ \bm{c}^\top \bm{z} \\
        %+ y^TPy+Q^Ty \\
        s.t. & \ \bm{z}\in \mathbf{Z}, \bm{A}\bm{z} + \bm{u} = \bm{0}, \\
        & \bm{u}_k \in  \mathbf{U}_k, \ \  \forall k \in \mathbf{K}.
    \end{split}
\end{equation}

\begin{remark}
    \textcolor{black}{The optimal dispatch model proposed above can be generalized to situations with other storage-like flexible resources integrated into the grid, like batteries or electric vehicles. Take the battery model as an example. We can regard $x_k^t$ as the state of charge of the battery and $u_k^t$ as the charging/discharging power and then set $M=1$.}
\end{remark}
\begin{remark}
    The optimal dispatch model proposed above is a centralized optimization with one coordinator, i.e., the DSO, and multiple agents, i.e., the BLAs. To construct \eqref{P0}, the BLA $k$ needs to upload the sensitive information, i.e., $\bm{R}_k$, $\bm{S}_k$, $\bm{d}_k$, $\bar{x}_k$ and $\underline{x}_k$ to the DSO. Besides, the DSO can infer the aggregate profile $\bm{x}_k$ by solving \eqref{P0}. These can bring serious privacy concerns to the BLA and occupants belonging to it.
\end{remark}

\vspace{-0.4cm}
\section{The Privacy-Preserved Algorithm}
\label{Section III}
This section proposes a privacy-preserved algorithm for \eqref{P0}. First, we provide the preliminaries for the privacy definition and the related privacy-preserved methods. Second,  we design a transformation-based encryption algorithm. Finally, we provide a detailed privacy analysis for the proposed algorithm.

\vspace{-0.3cm}
\subsection{Preliminaries}
\subsubsection{Privacy Definition}
In this work, the privacy of the BLA $k$ is defined as follows:
\begin{definition}
   For the BLA $k$ ($\forall k\in\mathbf{K})$, the privacy includes the parameter matrix $\bm{R}_k$, 
   %(i.e., $\alpha_k^m, \forall m \in \mathbf{M}\backslash\{0\}$), 
   the parameter matrix $\bm{S}_k$, 
   %(i.e., $\beta_k^m, \forall m \in\mathbf{M}$)
   the constant vector $\bm{d}_k$,
   %(i.e., $\alpha_k^m, \forall m \in \mathbf{M} \backslash \{0\}$, $\beta_k^m, \forall m\in \mathbf{M}$, and $\gamma_k^t$),  
   the upper and lower bounds of the state variables (i.e., $\bar{x}_k$ and $\underline{x}_k$), and the state variables $\bm{x}_k$.
   %Namely, the constant matrices $\bm{A}_k$, $\bm{B}_k$, $f$, and the state vector $y$ are private information that needs protection.
\end{definition}
\subsubsection{Transformation-Based Encryption Method}
The TE is a computationally efficient method with a strong privacy guarantee. We take the linear matrix equality as an example to illustrate the TE method, defined as 
\begin{subequations}
\begin{equation}
    \bm{B}\bm{y} = \bm{b}.
\end{equation}
wherein $\bm{y}\in\mathbb{R}^{T\times 1}$ is the decision vector, and $\bm{B}\in\mathbb{R}^{T\times T}$ and $\bm{b}\in\mathbb{R}^{T\times 1}$ are the coefficient matrices belonging to an agent. The computing center needs $\bm{B}$ and $\bm{b}$ to implement the optimization calculation, which causes private information leakage. The fundamental idea of the TE method is to conceal private information with random matrices. There are two ways to implement the TE method, i.e., random mapping and random multiplication, referred to as TE-I and TE-II in this paper. Typically, they need to be combined to protect privacy information.

(1) TE-I:
To protect the random matrix $\bm{B}$, the agent generates a random invertible matrix $\bm{W}\in\mathbb{R}^{T\times T}$ to implement the random mapping as follows:
\begin{equation}
    \bm{y} = \bm{W}\tilde{\bm{y}}.
\end{equation}
Then, the original constraints are transformed into 
\begin{equation}
    \bm{B}\bm{W}\tilde{\bm{y}}=\bm{b}.
\end{equation}
wherein the private matrix $\bm{B}$ is masked by the random matrix $\bm{W}$. It is worth mentioning that this transformation is bijective because of the invertibility of the random matrix $\bm{W}$. Namely, there exists a unique inverse transformation, i.e., $\tilde{\bm{y}}=\bm{W}^{-1}\bm{y}$.
However, The TE-I method has limitations: it cannot protect the private coefficients in the $\bm{y}$-unrelated terms, i.e., the vector $\bm{b}$ in the above constraint.

(2) TE-II: To protect the random vector $\bm{b}$, the agent generates a random invertible matrix $\bm{V}\in\mathbb{R}^{T\times T}$, multiplied on both sides of the linear matrix equality, as 
\begin{equation}
    \bm{V}\bm{B}\bm{W}\tilde{\bm{y}}=\bm{V}\bm{b}.
\end{equation}
Then, the agent can convey the masked matrices $\bm{V}\bm{B}\bm{W}$ and $\bm{V}\bm{b}$ to the computing center for optimization calculation. After receiving the optimal result $\tilde{\bm{y}}^*$ from the computing center, the agent can recover the original optimal result as 
\begin{equation}
    \bm{y}^* = \bm{W} \tilde{\bm{y}}^*.
\end{equation}
\end{subequations}

\vspace{-0.5cm}
\subsection{The Proposed Privacy-Preserved Algorithm}
\begin{figure}[t]
    \centering
    \includegraphics[width=0.9\columnwidth]{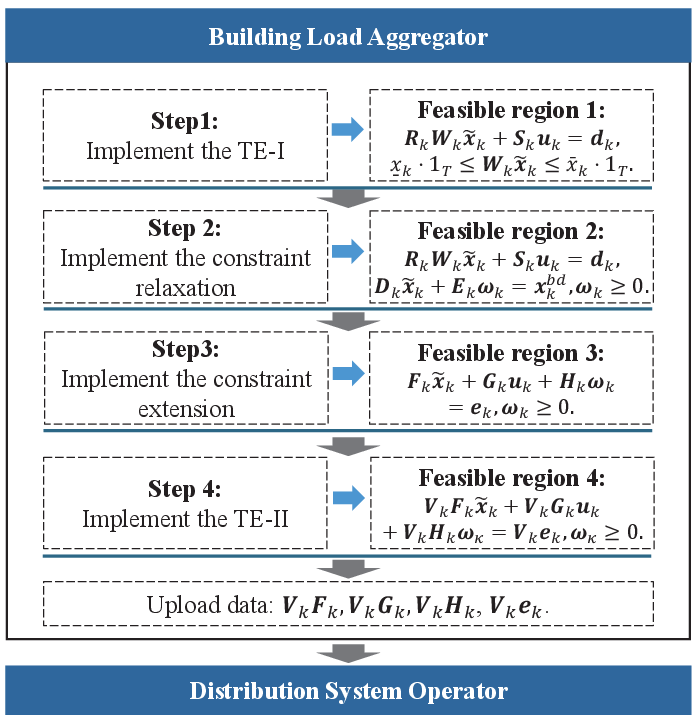}
    \setlength{\abovecaptionskip}{-5pt}
    \vspace{0.4cm}
    %\captionsetup{labelfont={color=blue}}
    \caption{\textcolor{black}{The information encryption mechanism and the equivalent transformation of the feasible region.}}
    \label{information encryption mechanism}
\end{figure}

To protect private information, the BLA $k$ cannot directly upload the parameter matrices $\bm{R}_k$ and $\bm{S}_k$, the vector $\bm{d}_k$, and the upper (or lower) bound information $\bar{x}_k$ (or $\underline{x}_k$) to the DSO. Thus, we design the privacy-preserved algorithm for the optimal dispatch model \eqref{P0}, the principles of which include the TE-I, the constraint relaxation technique, the constraint extension technique, and the TE-II. The information encryption mechanism is shown in Fig. \ref{information encryption mechanism}.
In the following, we introduce each part in detail.

\subsubsection{The Implementation of TE-I}
First, the BLA $k$ generates an invertible random matrix $\bm{W}_k\in\mathbb{R}^{T\times T}$, which is not disclosed to the DSO and other BLAs. Using this, the state variable $\bm{x}_k$ can be mapped into a pseudo-state, i.e., $\tilde{\bm{x}}_k$, as 
\begin{subequations}
\begin{equation}
    \label{TE-I}
    \tilde{\bm{x}}_k = \left(\bm{W}_k\right)^{-1} \bm{x}_k.
\end{equation}

Then, the model of the BLA $k$ can be equivalently transformed as 
\begin{equation}
    \label{BLA model2-1}
    \bm{R}_k\bm{W}_k \tilde{\bm{x}}_k+\bm{S}_k\bm{u}_k=\bm{d}_k,
\end{equation}
\begin{equation}
    \label{BLA model2-2}
    \underline{x}_k\cdot \bm{1}_T \leq \bm{W}_k \tilde{\bm{x}}_k\leq\bar{x}_k\cdot \bm{1}_T.
\end{equation}
\end{subequations}

However, this transformation is still not enough to protect the privacy of the BLA $k$ because it has to share $\bm{d}_k$, $\bm{S}_k$, $\bar{x}_k$, $\underline{x}_k$ and $\bm{W}_k$ with the DSO. In addition to the private information that can be obtained directly, i.e., $\bm{d}_k$, $\bm{S}_k$, $\bar{x}_k$ and $\underline{x}_k$, the DSO can also use the random matrix $\bm{W}_k$ to recover the real state information $\bm{x}_k$ according to the inverse mapping of \eqref{TE-I}. To solve this problem, the naive method is to apply the transformation TE-II to the BLA model \eqref{BLA model2-1}-\eqref{BLA model2-2}, namely, multiply on both sides of the constraints with random matrices. However, this can lead to privacy leakage, the reason for which will be detailed in Section \ref{The Constraint Relaxation Technique}.
\begin{remark}
Applying the TE-I to $\bm{u}_k$ has no contribution to privacy protection. The reason is as follows:
    We denote the coupling constraint related to the BLA $k$ in \eqref{P0} as $\bm{A}_k\bm{z}+\bm{u}_k=\bm{0}$. Assuming that the TE-I is applied to $\bm{u}_k$, i.e., $\bm{u}_k = \bm{W}_k^u \tilde{\bm{u}}_k$, the above coupling constraint is equivalently transformed as 
\begin{equation*}
    \label{appendix b1}
    \bm{A}_k\bm{z}+\bm{W}_k^u\tilde{\bm{u}}_k=\bm{0}.
\end{equation*}
To construct the optimal dispatch problem, the BLA $k$ needs to upload the coefficient matrix $\bm{W}_k^u$ to the DSO. However, once the random matrix $\bm{W}_k^u$ is known, using a random matrix to mask parameters, e.g., $\bm{S}_k$, becomes ineffective. Namely, the coupling constraint invalidates the effect of TE-I on $\bm{u}_k$.
\end{remark}

\subsubsection{The Implementation of the CRT}
\label{The Constraint Relaxation Technique}
In this part, we first illustrate why we transform inequality into equality before implementing the TE-II. Then, we transform the BLA model \eqref{BLA model2-1}-\eqref{BLA model2-2} into the equality constraints by introducing the slack variables. 

(1) The necessity of implementing the CRT:
 The constraint \eqref{BLA model2-2} is an inequality constraint, imposing stringent restrictions on the application of the TE-II to ensure the feasible region remains unchanged: the transformation matrix multiplied on both sides of the \eqref{BLA model2-2} should be a positive diagonal matrix. This significantly reduces the number of random variables, increasing the privacy leakage risk. 
To elaborate on this viewpoint, we consider the situation in which the BLA $k$ applies the TE-II without transforming the inequality into equality. The BLA $k$ generates two invertible random matrices $\bm{V}_k^1\in\mathbb{R}^{T\times T}$ and $\bm{V}_k^2\in\mathbb{R}^{2T\times 2T}$, and multiply them on both sides of \eqref{BLA model2-1} and \eqref{BLA model2-2}. Considering the constraint \eqref{BLA model2-2} is inequality, $\bm{V}_k^2$ must be a positive diagonal matrix to keep the feasible region unchanged. Thus, the BLA model is transformed into
\begin{subequations}
\label{BLA model-why need transformation from inequality into equality}
    \begin{equation}
    \label{BLA model-why need transformation from inequality into equality-1}
\bm{V}_k^1\bm{R}_k\bm{W}_k\tilde{\bm{x}}_k+\bm{V}_k^1\bm{S}_k\bm{u}_k=\bm{V}_k^1\bm{d}_k,
\end{equation}
\begin{equation}
        \label{BLA model-why need transformation from inequality into equality-2}
    \bm{V}_k^2\underline{x}_k\cdot \bm{1}_T \leq \bm{V}_k^2\bm{W}_k\tilde{\bm{x}}_k\leq
\bm{V}_k^2\bar{x}_k\cdot \bm{1}_T.
\end{equation}
\end{subequations}

Then, the BLA $k$ uploads $\bm{V}_k^1\bm{R}_k\bm{W}_k$, $\bm{V}_k^1\bm{S}_k$, $\bm{V}_k^1\bm{d}_k$, $\bm{V}_k^2\underline{x}_k\cdot \bm{1}_T$, $\bm{V}_k^2\bm{W}_k$ and $\bm{V}_k^2\bar{x}_k\cdot\bm{1}_T$ to the DSO. The number of inference equations they provide are $T^2$, $T^2$, $T$, $T$, $T^2$ and $T$ respectively, totally $3T^2+3T$ in number. On the other hand, the random matrices $\bm{V}_k^1$, $\bm{V}_k^2$, $\bm{W}_k$ provide $T^2$, $2T$ and $T^2$ unknown variables, respectively; the coefficient matrices $\bm{R}_k$, $\bm{S}_k$, $\bm{d}_k$, $\bar{x}_k$ and $\underline{x}_k$ provide $M$, $M+1$, $T$, $1$ and $1$ unknown variables. The total number of the unknown variables is $2T^2+3T+2M+3$. Considering the economic dispatch period, $T$ is much larger than the model order, $M$, and the number of unknown variables is less than that of inference equations. Therefore, the equation system is over-determined, and the DSO may get the solution. Namely, this method has a risk of leaking private information. Thus, we need to unify the BLA model \eqref{BLA model2-1}-\eqref{BLA model2-2} into the equality form.

(2) The implementation of the CRT:
First, we denote $\bm{D}_k=[\bm{W}_k^\top, -\bm{W}_k^\top]^\top\in\mathbb{R}^{2T\times T}$, $\bm{x}_k^{bd}=[(\bar{x}_k\cdot \bm{1}_T)^\top,(-\underline{x}_k\cdot \bm{1}_T)^\top]^\top\in\mathbb{R}^{2T\times 1}$ and introduce the slack vector $\bm{w}_k\in\mathbb{R}^{2T\times 1}$ ($\bm{w}_k\geq 0$). Then, the BLA model \eqref{BLA model2-1}-\eqref{BLA model2-2} is equivalently transformed as 
\begin{subequations}
\begin{equation}
    \label{BLA model3-1}\bm{R}_k\bm{W}_k\tilde{\bm{x}}_k+\bm{S}_k\bm{u}_k=\bm{d}_k,
\end{equation}
\begin{equation}
    \label{BLA model3-2}
    \bm{D}_k\tilde{\bm{x}}_k+\bm{E}_k\bm{w}_k= \bm{x}_k^{bd}, \ \bm{w}_k\geq 0.
\end{equation}
\end{subequations}
It is worth mentioning that $\bm{E}_k\in\mathbb{R}^{2T\times 2T}$ should be a random diagonal matrix with positive diagonal entries and privately owned by the BLA $k$. Introducing  $\bm{E}_k$ can enhance privacy protection while guaranteeing the equivalence between \eqref{BLA model2-2} and \eqref{BLA model3-2}.  The above transformation from inequality to equality lays a foundation for applying the TE-II. However, if we directly apply the TE-II to the constraint \eqref{BLA model3-1}-\eqref{BLA model3-2}, there is still a risk of privacy leakage, the reason for which will be elaborated in Section \ref{The Constraint Extension Technique}.

\subsubsection{The Implementation of the CET}
\label{The Constraint Extension Technique}
 In this part, we first elaborate on the necessity of introducing the CET before applying the TE-II and then explain its implementation.

(1) The necessity of implementing the CET:
Consider that we directly apply the TE-II to \eqref{BLA model3-1}-\eqref{BLA model3-2} without implementing the CET. In this case, the BLA $k$ first reformulates the BLA model \eqref{BLA model3-1}-\eqref{BLA model3-2} into 
\begin{equation}
    \label{BLA model-appendix_d-1}
\bm{F}_k’\tilde{\bm{x}}_k+\bm{G}_k'\bm{u}_k+\bm{H}_k'\bm{w}_k = \bm{e}_k', \ \bm{w}_k\geq 0,
\end{equation}
wherein $\bm{F}_k'$, $\bm{G}_k'$, $\bm{H}_k'$ and $\bm{e}_k'$ are presented as \eqref{F}-\eqref{e}. We use the 'prime' symbol to distinguish \eqref{BLA model-appendix_d-1} and \eqref{BLA model4}.
\begin{subequations}
    \label{FGHe'}
 \begin{equation}
     \bm{F}_k' = \left[
 \begin{array}{c}
    \bm{R}_k\bm{W}_k \\
    \bm{D}_k \\
\end{array}
\right]\in\mathbb{R}^{3T\times T},
 \end{equation}
\begin{equation}
    \bm{G}_k' = \left[
\begin{array}{cccc}
     \bm{S}_k^\top &  \bm{0}_{T\times 2T}\\
\end{array}
\right]^\top\in\mathbb{R}^{3T\times T},
\end{equation}
\begin{equation}
    \bm{H}_k' = \left[
\begin{array}{cccc}
     \bm{0}_{2T\times T} & \bm{E}_k^\top\\
\end{array}
\right]^\top\in\mathbb{R}^{3T\times 2T},
\end{equation}
\begin{equation}
    \bm{e}_k' = \left[
\begin{array}{cccc}
     \bm{d}_k^\top  & \bm{x}_k^{bd\top}\\
\end{array}
\right]^\top\in\mathbb{R}^{3T\times 1}.
\end{equation}
 \end{subequations}

Then, We multiply on both sides of the constraint $\eqref{BLA model-appendix_d-1}$ with the random matrix $\bm{V}_k'\in\mathbb{R}^{3T\times 3T}$, wherein the matrix $\bm{V}_k'$ is generated and privately owned by the BLA $k$, as 
\begin{equation}
    \label{BLA model-appendix_d-2}
\bm{V}_k'\bm{F}_k'\tilde{\bm{x}}_k+\bm{V}_k'\bm{G}_k'\bm{u}_k+\bm{V}_k'\bm{H}_k'\bm{w}_k = \bm{V}_k'\bm{e}_k', \ \bm{w}_k\geq 0.
\end{equation}

Then, the BLA $k$ needs to upload $\bm{V}_k'\bm{F}_k'$, $\bm{V}_k'\bm{G}_k'$, $\bm{V}_k'\bm{H}_k'$ and $\bm{V}_k'\bm{e}_k'$ to the DSO, which provides $3T^2$, $3T^2$, $6T^2$ and $3T$ inference equations respectively, totally $12T^2+3T$ in number. The matrices $\bm{V}_k'$, $\bm{W}_k
$, $\bm{E}_k$, $\bm{R}_k$, $\bm{S}_k$, $\bm{d}_k$ and $\bm{x}_k^{bd}$ provides $9T^2$, $T^2$, $2T$, $M$, $M+1$, $T$ and $2$ respectively, totally $10T^2+3T+2M+3$ in number. It is obvious that when $T\geq M+2$ (practically easy to satisfy), the total number of inference equations is larger than that of the unknown variables, making \eqref{BLA model-appendix_d-2} an over-determined system. This arouses potential privacy leakage of the BLA $k$. Thus, we propose the CET to pre-process the BLA model \eqref{BLA model3-1}-\eqref{BLA model3-2} before implementing the TE-II in the following.

(2) The implementation of the CET:
We first duplicate the BLA model \eqref{BLA model3-1}-\eqref{BLA model3-2} and reformulate them as 
\begin{subequations}
\begin{equation}
    \label{BLA model4}
    \bm{F}_k\tilde{\bm{x}}_k+\bm{G}_k\bm{u}_k+\bm{H}_k\bm{w}_k = \bm{e}_k, \ \bm{w}_k\geq 0,
\end{equation}
wherein the matrices $\bm{F}_k$, $\bm{G}_k$, $\bm{H}_k$, and $\bm{e}_k$ are defined as
\label{FGHe}
 \begin{equation}
 \label{F}
     \bm{F}_k = \left[
 \begin{array}{c}
    \bm{R}_k\bm{W}_k \\
    \bm{R}_k\bm{W}_k \\
    \bm{D}_k \\
    \bm{D}_k \\
\end{array}
\right]\in\mathbb{R}^{6T\times T},
 \end{equation}
\begin{equation}
\label{G}
    \bm{G}_k = \left[
\begin{array}{cccc}
     \bm{S}_k^\top & \bm{S}_k^\top & \bm{0}_{T\times 2T} & \bm{0}_{T\times 2T}\\
\end{array}
\right]^\top\in\mathbb{R}^{6T\times T},
\end{equation}
\begin{equation}
\label{H}
    \bm{H}_k = \left[
\begin{array}{cccc}
     \bm{0}_{2T\times T} & \bm{0}_{2T\times T} & \bm{E}_k^\top & \bm{E}_k^\top\\
\end{array}
\right]^\top\in\mathbb{R}^{6T\times 2T},
\end{equation}
\begin{equation}
\label{e}
    \bm{e}_k = \left[
\begin{array}{cccc}
     \bm{d}_k^\top & \bm{d}_k^\top & \bm{x}_k^{bd\top} & \bm{x}_k^{bd\top}\\
\end{array}
\right]^\top\in\mathbb{R}^{6T\times 1}.
\end{equation}
\end{subequations}
\begin{remark}
    It is worth mentioning that while we duplicate the constraints only once in this paper, they can be duplicated various times tailored to different problems. The more times the constraints are duplicated, the higher the dimension of the random matrix $\bm{V}_k$ is, which enhances the privacy protection performance but increases computational complexity.
\end{remark}

%\vspace{-0.1cm}
\subsubsection{The Implementation of TE-II}
\label{the Implementation of TE-II}
After implementing the CET, the BLA $k$ can apply the TE-II to \eqref{BLA model4}. Specifically, a random invertible matrix $\bm{V}_k\in\mathbb{R}^{6T\times 6T}$, privately owned by the BLA $k$, is generated and multiplied on both sides of the BLA model \eqref{BLA model4}, as 
\begin{equation}
    \label{BLA model5}
    \bm{V}_k\bm{F}_k\tilde{\bm{x}}_k+\bm{V}_k\bm{G}_k\bm{u}_k+\bm{V}_k\bm{H}_k\bm{w}_k = \bm{V}_k\bm{e}_k.
\end{equation}
Then, the BLA $k$ needs to upload $\bm{V}_k\bm{F}_k$, $\bm{V}_k\bm{G}_k$, $\bm{V}_k\bm{H}_k$ and $\bm{V}_k\bm{e}_k$ to the DSO for optimization calculation. Define $\bm{w} \triangleq  \left[ \bm{w}_1^\top,\cdots,\bm{w}_{K}^\top \right ]^\top\in\mathbb{R}^{2TK\times 1}$, Define $\tilde{\bm{x}} \triangleq  \left[ \tilde{\bm{x}}_1^\top,\cdots,\tilde{\bm{x}}_{K}^\top \right ]^\top\in\mathbb{R}^{TK\times 1}$ The masked optimization problem for the DSO is as \text{P1}:
\begin{equation}
    \label{P1}
     \begin{split}
        \text{P1}: \ \underset{\bm{z},\tilde{\bm{x}},\bm{u},\bm{w}}{\min} & \ \bm{c}^\top \bm{z} \\
        %+ y^TPy+Q^Ty \\
        s.t. & \ \bm{z}\in \mathbf{Z}, \bm{A}\bm{z} + \bm{u} = \bm{0}, \\
        & \bm{u}_k \in  \mathbf{U}_k, \ \  \forall k \in \mathbf{K},
    \end{split}
\end{equation}
wherein the feasible region of $\bm{u}_k$ is represented as
\begin{equation*}
\begin{split}
\mathbf{U}_k = \{ &\bm{u}_k|
\bm{V}_k\bm{F}_k\tilde{\bm{x}}_k+\bm{V}_k\bm{G}_k\bm{u}_k+\bm{V}_k\bm{H}_k\bm{w}_k = \bm{V}_k\bm{e}_k\}.
\end{split}
\end{equation*}

\begin{figure}[t]
    \centering
    \includegraphics[width=\columnwidth]{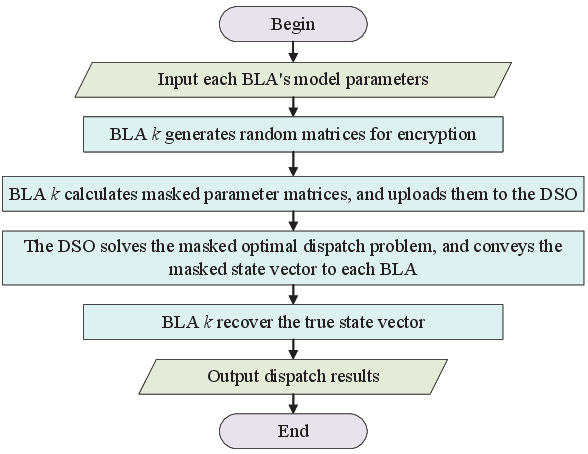}
    \setlength{\abovecaptionskip}{-5pt}
    \vspace{0.2cm}
    %\captionsetup{labelfont={color=blue}}
    \caption{\textcolor{black}{The flowchart of the proposed privacy-preserved optimal dispatch algorithm.}}
    \label{privacy-preserved optimal dispatch algorithm}
\end{figure}

\textcolor{black}{The flowchart that combines the dispatch model with the privacy-preserved algorithm is presented in Fig. \ref{privacy-preserved optimal dispatch algorithm}. 
The whole privacy-preserved algorithm is presented in Algorithm \ref{Algoritm 1}. }

%For convenience, we denote $\bm{V}_k\bm{F}_k$, $\bm{V}_k\bm{G}_k$, $\bm{V}_k\bm{H}_k$ and $\bm{V}_k\bm{e}_k$ as $\bm{f}_k^1$, $\bm{f}_k^2$, $\bm{f}_k^3$ and $\bm{f}_k^4$ respectively.

\begin{algorithm}[t]
\small
    \SetAlgoLined %显示end
	\caption{Proposed privacy-preserved algorithm.} 
         \label{Algoritm 1}
        The BLA $k$ generates the random invertible matrices $\bm{W}_k$ and $\bm{E}_k$, and formulate the masked matrices $\bm{F}_k$, $\bm{G}_k$, $\bm{H}_k$ and $\bm{e}_k$ according to \eqref{F}-\eqref{e}\;
        The BLA $k$ generates the random invertible matrix $\bm{V}_k$ and formulate the matrices $\bm{V}_k\bm{F}_k$, $\bm{V}_k\bm{G}_k$, $\bm{V}_k\bm{H}_k$ and $\bm{V}_k\bm{e}_k$\;
        The BLA $k$ uploads $\bm{V}_k\bm{F}_k$, $\bm{V}_k\bm{G}_k$, $\bm{V}_k\bm{H}_k$ and $\bm{V}_k\bm{e}_k$ to the DSO\;
        The DSO formulates the optimal dispatch model \eqref{P1} and solves it\;
        The DSO distributes the masked solution of state, i.e., $\tilde{\bm{x}}_k$ to the BLA $k$\;
        The BLA $k$ recover the real state $\bm{x}_k$ according to $\bm{x}_k =\bm{W}_k\tilde{\bm{x}}_k$.
\end{algorithm}

\vspace{-0.3cm}
\subsection{Privacy Analysis}
\label{privacy analysis}
This section conducts a detailed privacy analysis for our proposed computation algorithm. \textcolor{black}{For a specific BLA, we will consider potential privacy inference from semi-honest adversaries, i.e., the DSO and other BLAs, as well as privacy threats from external eavesdroppers.}

\subsubsection{\textcolor{black}{Semi-honest adversaries}}
\textcolor{black}{As for the privacy protection issue of a specific BLA, two semi-honest adversaries, i.e., the DSO and other BLAs are considered in this paper. }

\textcolor{black}{First, we investigate the potential for the DSO to infer privacy information of BLA $k$.} The DSO can gather received information to construct the multivariate inference equation system. 
%The core of realizing privacy protection is to ensure the number of inference equations is less than that of unknown variables, no matter what information the DSO selects to construct the inference equation system.
Denote $\bm{f}_k^1\triangleq\bm{V}_k\bm{F}_k\in\mathbb{R}^{6T\times T}$, $\bm{f}_k^2\triangleq\bm{V}_k\bm{G}_k\in\mathbb{R}^{6T\times T}$, $\bm{f}_k^3\triangleq\bm{V}_k\bm{H}_k\in\mathbb{R}^{6T\times 2T}$ and $\bm{f}_k^4\triangleq\bm{V}_k\bm{e}_k\in\mathbb{R}^{6T\times 1}$. When the DSO receive the information uploaded by the BLA $k$, $\bm{f}_k^1$, $\bm{f}_k^2$, $\bm{f}_k^3$ and $\bm{f}_k^4$ become known information while $\bm{V}_k$, $\bm{F}_k$, $\bm{G}_k$, $\bm{H}_k$ and $\bm{e}_k$ are the matrices to be inferred. The complete inference equation system for BLA $k$ can be formulated as
\begin{subequations}
    \begin{equation}
        \label{inference equation-1}
        \bm{V}_k\bm{F}_k = \bm{f}_k^1,
    \end{equation}
    \begin{equation}
        \label{inference equation-2}
        \bm{V}_k\bm{G}_k = \bm{f}_k^2,
    \end{equation}
    \begin{equation}
        \label{inference equation-3}
        \bm{V}_k\bm{H}_k = \bm{f}_k^3,
    \end{equation}
    \begin{equation}
        \label{inference equation-4}
        \bm{V}_k\bm{e}_k = \bm{f}_k^4,
    \end{equation}
\end{subequations}
According to the dimension of $\bm{f}_k^i$ ($i=1,\cdots,4$), \eqref{inference equation-1}-\eqref{inference equation-4} provide $6T^2$, $6T^2$, $12T^2$ and $6T$ inference equations respectively. On the other hand, the random matrix $\bm{V}_k$ has $6T\times 6T=36T^2$ unknown variables. We note that the number of unknown variables $\bm{V}_k$ provides is larger than the total number of the inference equations, i.e., $24T^2+6T$. This means the equation system is under-determined. Namely, neither the random matrices $\bm{W}_k$ and $\bm{V}_k$ or the coefficient matrices $\bm{F}_k$, $\bm{G}_k$, $\bm{H}_k$ and $\bm{e}_k$ can be uniquely determined. Thus, the private information $\bm{R}_k$, $\bm{S}_k$, $\bm{d}_k$, $\bar{x}_k$ and  $\underline{x}_k$ are protected. Besides, the DSO cannot infer the real state of BLA according to $\bm{x}_k=\bm{W}_k\tilde{\bm{x}}_k$ if not knowing $\bm{W}_k$. In summary, the DSO cannot infer the private information of the BLA $k$.

\textcolor{black}{Second, we state that the information accessible to other BLAs is a subset of that available to the DSO (even if they collude with each other). Since the DSOs cannot infer private information of the BLA $k$, no other BLAs can achieve privacy inference.}

\subsubsection{\textcolor{black}{External eavesdroppers.}}
\textcolor{black}{Then, external eavesdroppers who may steal data transferred between the DSO and BLAs are considered. Because all the transferred data, i.e., $\bm{V}_k\bm{F}_k$, $\bm{V}_k\bm{G}_k$, $\bm{V}_k\bm{H}_k$ and $\bm{V}_k\bm{e}_k$,
are masked by random matrices, these adversaries cannot directly steal useful information. Besides, the information accessible to external eavesdroppers constitutes a subset of the DSO's available information. Given the DSO's inability to make privacy inferences from this limited dataset, these adversaries' capacity for inferring privacy is even more constrained.}

\begin{remark}
    \textcolor{black}{The proposed approach operates under the assumption of a secure communication environment, thus excluding potential threats like noise injection attacks that may compromise data integrity.}
\end{remark}
\section{Numerical Tests}
\label{Section IV}
\begin{figure}[t]
    \centering
    \includegraphics[width=1\columnwidth]{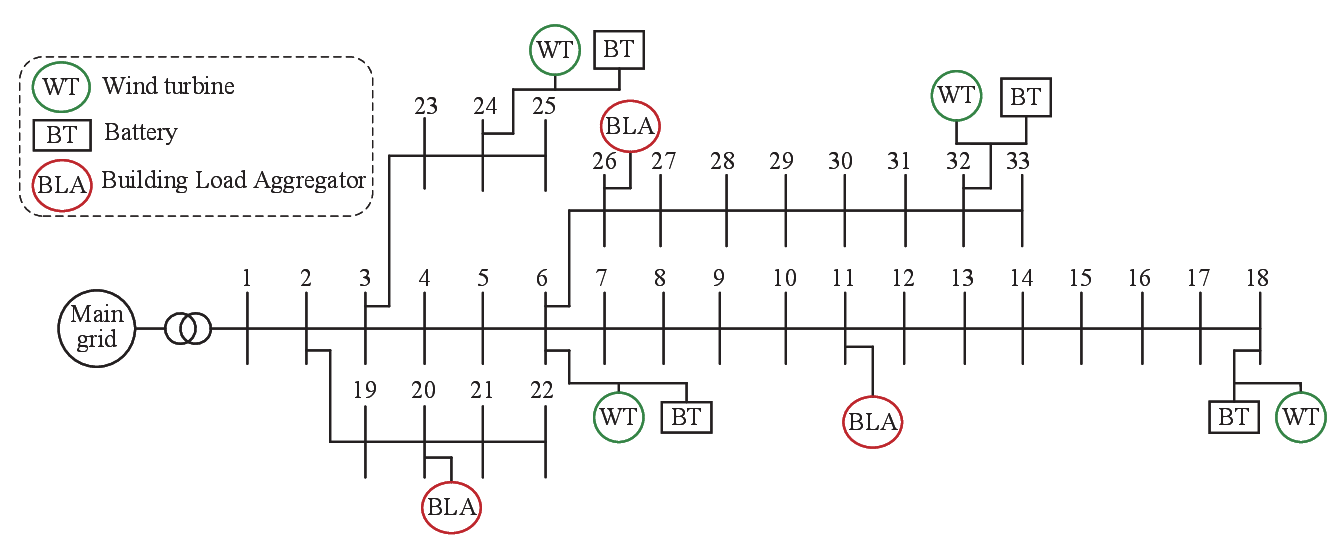}
    \setlength{\abovecaptionskip}{-5pt}
    \vspace{-0.2cm}
    %\captionsetup{labelfont={color=blue}}
    \caption{\textcolor{black}{The modifies IEEE 33 bus distribution system structure adopted in this paper.}}
    \label{system structure}
\end{figure}

To verify the effectiveness of the proposed privacy-preserved algorithm, we carry out numerical tests based on the modified IEEE 33 bus distribution system, as shown in Fig. \ref{system structure}. 
All the simulations are performed on a PC with an Intel i7 core and 32GB RAM. The programming is based on MATLAB R2022b and Yalmip. Gurobi 10.0.2 is used to solve the optimization problem. \textcolor{black}{The solver's relative MIP optimality gap is set as $10^{-6}$}. The model order of the BLA, $M$, is set as $1$. The initial aggregate temperature and injected power of each BLA are set as $23$ ℃ and $100$ kV, respectively. The parameter settings of the BLAs are shown in Table \ref{table: Parameter settings of BLAs}. \textcolor{black}{Other configurations of the IEEE 33 bus system are provided in \cite{Zeyin2025Parameters}.} We assume the non-zero elements in $\bm{W}_k$, $\bm{E}_k$ and $\bm{V}_k$ follow the Gaussian distribution $N(0.1, 0.1)$.

\begin{table}[t]
    \centering
    \footnotesize
    \caption{\textcolor{black}{Parameter settings of BLAs}}
    \label{table: Parameter settings of BLAs}
    %\vspace{-0.2cm}
    \resizebox{.9\columnwidth}{!}{
    \begin{tabular}{cccccccc}
    \toprule
    \textbf{No. of BLA} & $\alpha_k^0$ & $\alpha_k^1$ & $\beta_k^0$ & $\beta_k^1$ & $\gamma_k^t$ & $\bar{\tau}_k$ & $\underline{\tau}_k$ \\
    \midrule
    \textbf{1} & 1 & 0.96 & 0.005 & 0.003 & 0.02 & 27 & 23 \\
    \textbf{2} & 1 & 0.963 & 0.0051 & 0.002  & 0.01 & 26 & 22 \\
    \textbf{3} & 1 & 0.959 & 0.0056 & 0.004 & 0.02 & 26 & 22\\
    \bottomrule
\end{tabular}
}
\end{table}

% \subsection{Performance of the Proposed Algorithm}
%\subsection{Accuracy Analysis}
\subsection{Accuracy Analysis}
\begin{figure}[t]  
 \captionsetup{singlelinecheck=on}
  \footnotesize % 8pt
  \centering   
  \subfloat[]  
  {
    \label{fig:accuracy_1}\includegraphics[width=1\linewidth]{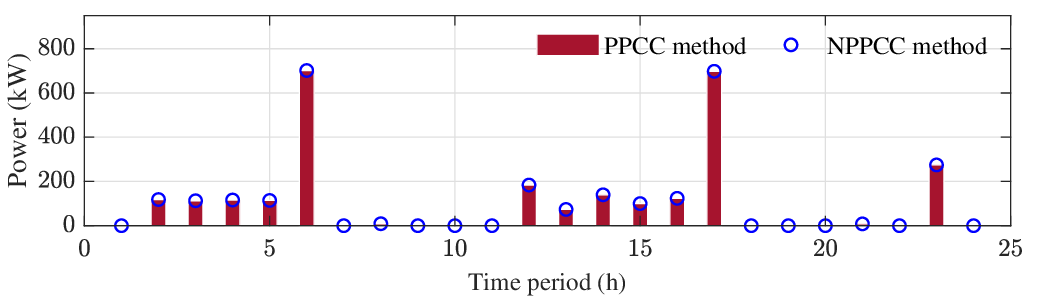}
  }\\
    \vspace{-0.35cm}
  \subfloat[] 
  {
\label{fig:accuracy_2}\includegraphics[width=1\linewidth]{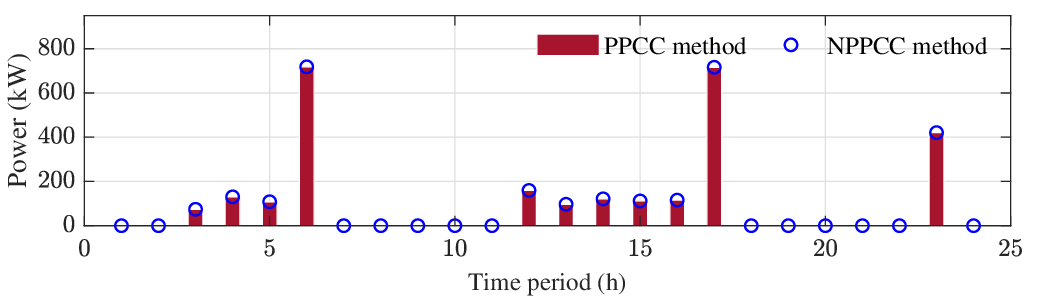}
  }\\   
 \vspace{-0.35cm}
  \subfloat[]
  {
 \label{fig:accuracy_3}\includegraphics[width=1\linewidth]{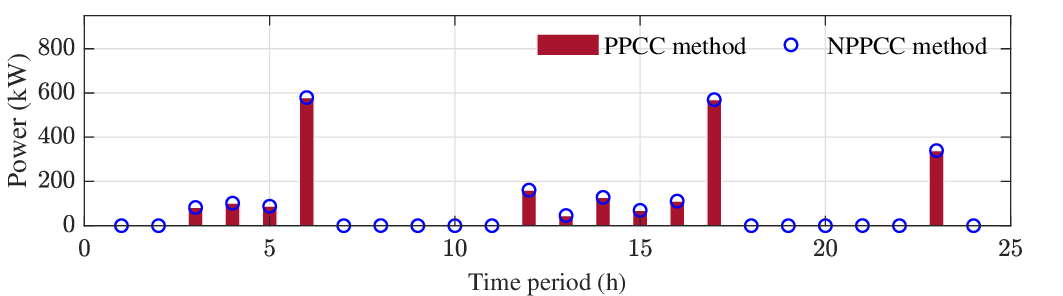}  
 }
  %\captionsetup{labelfont={color=blue}}
  \vspace{-0.1cm}
  \caption{\textcolor{black}{The injection power into BLAs, i.e., $\bm{u}_k$: (a) BLA 1; (b) BLA 2 (c) BLA 3.}}
  \label{fig:accuracy}  % 得放在最后，不然没法交叉引用
\end{figure}

\textcolor{black}{The centralized computing method without incorporating the privacy-preserved mechanism proposed in this paper is referred to as the non-privacy-preserved centralized computing (NPPCC), while our proposed one is termed as the privacy-preserved centralized computing (PPCC)}.
We compare the injection power to the BLAs calculated by the PPCC and NPPCC methods. As shown in Fig \ref{fig:accuracy}, the results obtained by the two algorithms are the same. Besides, the total costs of the distribution system calculated by the two methods are both $4.2097\times 10^4$ RMB. These indicate that the privacy-preserved algorithm does not compromise the accuracy of the optimal dispatch.

\textcolor{black}{Moreover, we compare the proposed algorithm with the privacy-preserved distributed computing (abbreviated as PPDC in this paper) method proposed in \cite{huang2019dp}. Applying the PPDC method, we use the alternating direction method of multiplier (ADMM) for distributed computing, and inject random noise with decaying amplitude into the exchanged variables in each iteration, as
\begin{equation*}
    \tilde{\bm{u}}_k=\bm{u}_k+ |\varphi|^l\bm{w}_k,
\end{equation*}
wherein $\tilde{\bm{u}}_k$ is the exchanged information between the DSO and the BLA $k$ (after being encrypted by random noise), $\varphi\in(0,1)$ is the decaying factor, $l$ is the iteration index, and $\bm{w}_k$ is the random noise injected to the original exchanged variables $\bm{u}_k$. Each element in the random vector, $\bm{w}_k$, is set to obey the Gaussian distribution $N(0,10)$, and the penalty factor $\rho$ in the ADMM algorithm is set to 0.01. We define the optimality loss as 
\begin{equation*}
    \text{Optimality loss} = \vert\frac{C^{PPDC}-C^{PPCC}}{C^{PPCC}}\vert \times 100\%,
\end{equation*}
wherein $C^{PPCC}$ and $C^{PPDC}$ are the operational costs calculated under the PPCC and PPDC methods respectively.
As shown in Table \ref{table: Operational cost and optimality loss under PPDC}, the PPDC method would cause at least $0.3088\%$ optimality loss ($\varphi=0$). Besides, as the increase of the decaying factor $\varphi$, the system operational cost and the optimality loss also significantly heighten. 
}

\textcolor{black}{
In summary, the privacy-preserved centralized computing method proposed in this paper brings no optimality loss, which is unattainable by the privacy-preserved distributed computing method.
}

\begin{table}[t]
    \centering
    \footnotesize
    \caption{\textcolor{black}{Operational cost and optimality loss under PPDC}}
    \label{table: Operational cost and optimality loss under PPDC}
    %\vspace{-0.2cm}
    \resizebox{1\columnwidth}{!}{
    \begin{tabular}{cccccc}
    \toprule
   $\varphi$ & 0 & 0.2 & 0.4 & 0.6 & 0.8 \\
      \midrule
   Operational cost (RMB) & 42227 & 42259 & 42288 & 42314 & 42348 \\
   Optimality loss (\%) & 0.3088 & 0.3848 & 0.4537 & 0.5155 & 0.5962  \\
     \bottomrule
\end{tabular}
}
\end{table}

\textcolor{black}{}

\subsection{Privacy Security Analysis}
\begin{figure}[t]  
 \captionsetup{singlelinecheck=on}
  \footnotesize % 8pt
  \centering   
  \subfloat[]  
  {
    \label{fig:heatmap_G_BLA1}\includegraphics[width=0.48\linewidth]{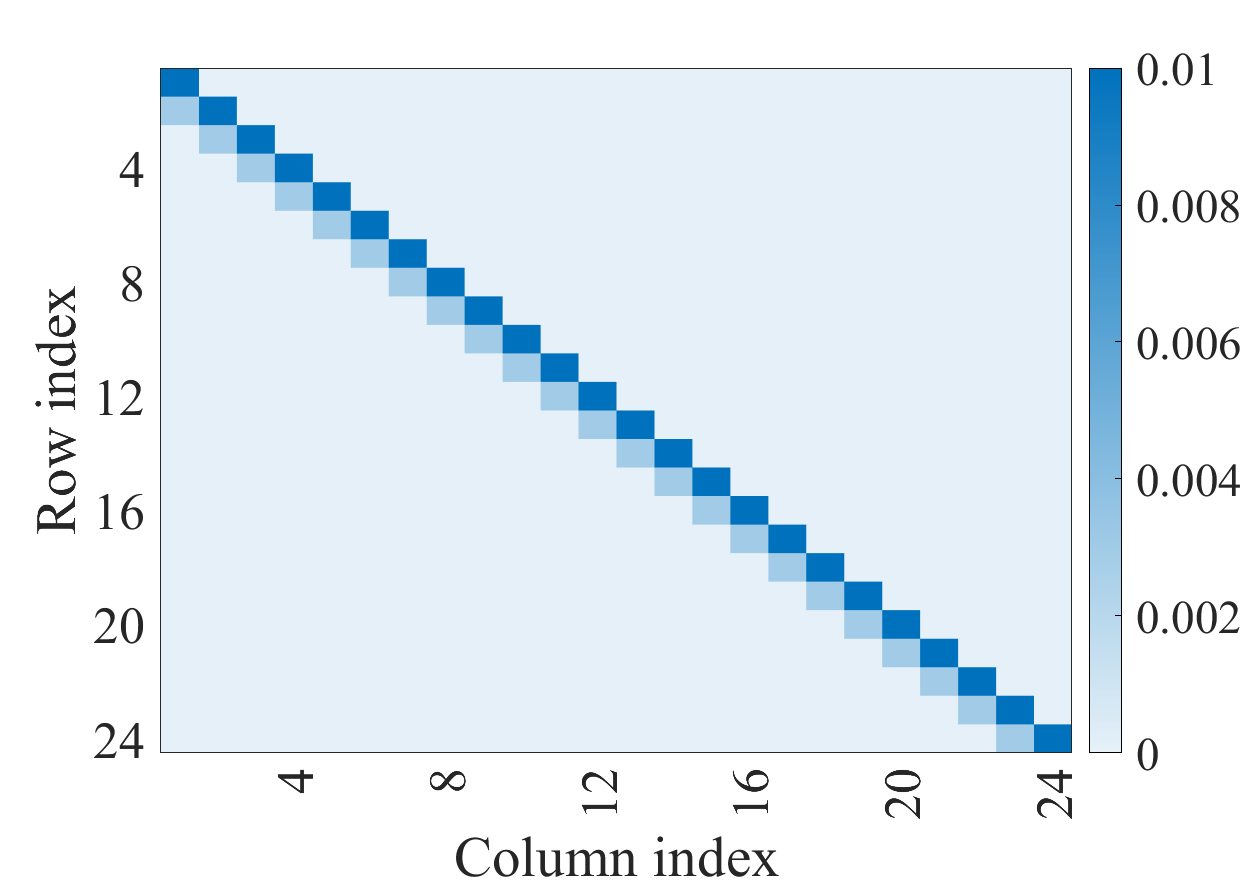}
  }
  \subfloat[] 
  {
\label{fig:heatmap_VG_BLA1}\includegraphics[width=0.48\linewidth]{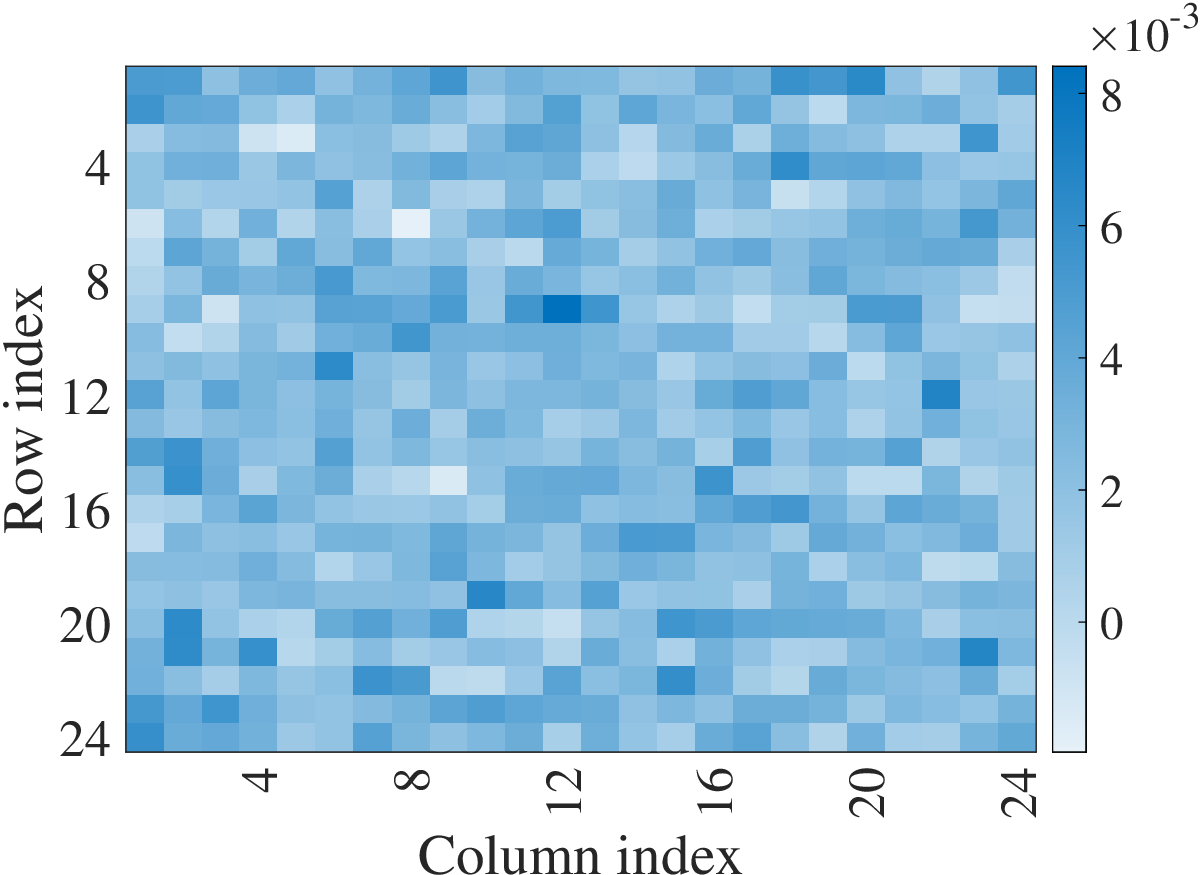}
  } 
  %\captionsetup{labelfont={color=blue}}
  \vspace{-0.1cm}
  \caption{The heatmaps of the first 24 rows of the matrices $\bm{G}$ and $\bm{V}_1\bm{G}_1$: (a) the matrix $\bm{G}_1$; (b) the matrix $\bm{V}_1\bm{G}_1$.}
  \label{fig:heatmap_G&VG_BLA1}  % 得放在最后，不然没法交叉引用
\end{figure}

To measure the performance of masking information, we compare the original data and masked data by the random matrix. We take the matrix $\bm{G}_1$, which contains the private information $\bm{S}_1$, as examples to illustrate the privacy-preserved performance. The heat maps in Fig \ref{fig:heatmap_G&VG_BLA1} can explain the great differences between the original matrices $\bm{G}_1$ and the masked matrix $\bm{V}_1\bm{G}_1$. Obviously, the sensitives information $\bm{G}_1$ is well masked by the random matrix $\bm{V}_1$.
% \subsection{Dispatch Result Analysis}

\subsection{\textcolor{black}{Influence of Operational Flexibility}}
\textcolor{black}{We assume that when a BLA's upper and lower temperature limits are equal, this BLA exhibits no flexibility and is therefore unable to contribute to regulation. In this situation, we denote $\bar{\tau}_k=\underline{\tau}_k=\tau_k^{const}$. Otherwise, we assume the BLA follows the default parameter settings as Table \ref{table: Parameter settings of BLAs}. We set the following four cases for comparison.}

\textcolor{black}{\textit{Case 1}: All three BLAs (1, 2, and 3) participate in regulation.}

\textcolor{black}{\textit{Case 2}: Only BLA 1 and BLA 2 participate in regulation; $\bar{\tau}_3=\underline{\tau}_3=\tau_3^{const}$.}

\textcolor{black}{\textit{Case 3}: Only BLA 1 participates in regulation; $\bar{\tau}_2=\underline{\tau}_2=\tau_2^{const}$; $\bar{\tau}_3=\underline{\tau}_3=\tau_3^{const}$.}

\textcolor{black}{\textit{Case 4}: None of the three BLAs (1, 2, or 3) participate in regulation; $\bar{\tau}_1=\underline{\tau}_1=\tau_1^{const}$; $\bar{\tau}_2=\underline{\tau}_2=\tau_2^{const}$; $\bar{\tau}_3=\underline{\tau}_3=\tau_3^{const}$.}

\textcolor{black}{The operational cost under different cases (\textit{Case 1} to \textit{Case 4}) and different temperature settings ($\tau_k^{const}$) is presented in Table \ref{table: operational cost}. (It is assumed that $\tau_1^{const}=\tau_2^{const}=\tau_3^{const}$ for simplicity.)
Obviously, as the increase in the number of BLAs engaged in system regulation, the operational cost of distribution system is significantly decrease. This demonstrates that BLAs possess flexible regulation capabilities, which effectively reduce the system operational cost.
Besides, elevating the fixed temperature setpoints of BLAs ($\tau_k^{const}$) escalates the system operational cost, attributable to the increased electricity demand required to sustain elevated temperature levels.}

\begin{figure}[t]
    \centering
    \includegraphics[width=1\columnwidth]{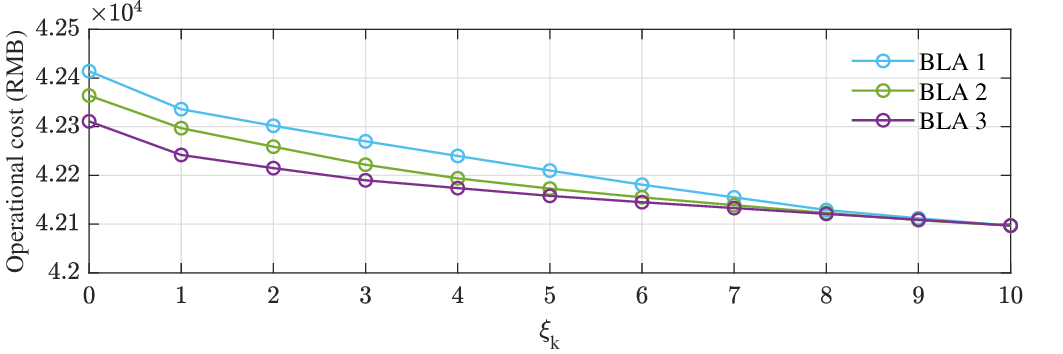}
    \setlength{\abovecaptionskip}{-5pt}
    \vspace{0.2cm}
    %\captionsetup{labelfont={color=blue}}
    \caption{\textcolor{black}{Operational cost under different permissible temperature range.}}
\label{fig: operational cost under different temperature range}
\end{figure}

\textcolor{black}{We denote the temperature limits $\bar{\tau}_k$ and $\underline{\tau}_k$ as:}
\begin{equation*}
    \bar{\tau}_k=\tau_k^{center}+\xi_k\Delta\tau_k,
\end{equation*}
\begin{equation*}
    \underline{\tau}_k=\tau_k^{center}-\xi_k\Delta\tau_k,
\end{equation*}
\textcolor{black}{
We set $\tau_1^{center}=25$, $\tau_2^{center}=\tau_3^{center}=24$, and $\Delta\tau_k=0.2$. By changing $\xi_k$ of an individual BLA and keeping the parameters of others as default, we obtain Fig. \ref{fig: operational cost under different temperature range}, which describes the relationship between the operational cost and the temperature range of one individual BLA. Fig. \ref{fig: operational cost under different temperature range} illustrates that as the permissible temperature range of the BLA expands, the system operational costs gradually decrease.}
\begin{table}[t]
    \centering
    \footnotesize
    \caption{\textcolor{black}{Operational cost under different cases and $\tau_k^{const}$.}}
    \label{table: operational cost}
    %\vspace{-0.2cm}
    \resizebox{0.97\columnwidth}{!}{
    \begin{tabular*}{\linewidth}{ccccc}
    \toprule
    \makebox[0.15\linewidth][c]{\textbf{$\bm{\tau}_k^{const}$}} &  \makebox[0.15\linewidth][c]{\textbf{Case 1}} & \makebox[0.15\linewidth][c]{\textbf{Case 2}} & \makebox[0.15\linewidth][c]{\textbf{Case 3}} & \makebox[0.15\linewidth][c]{\textbf{Case 4}}\\
    \midrule
    {23.5} & 42097 & 42262 & 42466 & 42609\\
    {24} & 42097 & 42311 & 42578 & 42779 \\
    {24.5} & 42097 & 42361 & 42690 & 42949 \\
    \bottomrule
\end{tabular*}
}
\end{table}

\subsection{Computational Performance Analysis}
\textcolor{black}{We implement the proposed privacy-preserved optimal dispatch algorithm in the above IEEE 33 bus and IEEE 141 bus distribution systems separately. The  configurations of both systems are elaborated in \cite{Zeyin2025Parameters}.}

\textcolor{black}{As shown in Table \ref{table: computational time (IEEE-33 cases)} and Table \ref{table: computational time (IEEE-141 cases)}, the computation time of the privacy-preserved algorithm is slightly longer than that of the non-privacy-preserved algorithm, regardless of the system size. This occurs because the proposed privacy-preserved algorithm introduces slack variables, i.e., $\bm{w}_k$, and duplicated constraints, thereby increasing computational complexity. It illustrates a trade-off between the privacy preservation effect and the computation efficiency. It is worth noting that the computational duration of this privacy-preserved algorithm remains low and does not significantly increase compared with the non-privacy-preserved one.}

\textcolor{black}{Besides, a comparative analysis of Table \ref{table: computational time (IEEE-33 cases)} and Table \ref{table: computational time (IEEE-141 cases)} reveals that the proposed method exhibits a significant increase in computation time when applied to larger system scale, yet remains within an acceptable range. This demonstrates the scalability and applicability of the proposed method.
}

\textcolor{black}{}
\begin{table}[t]
    \centering
    \footnotesize
    \caption{\textcolor{black}{computational time (IEEE-33 cases)}}
    \label{table: computational time (IEEE-33 cases)}
    %\vspace{-0.2cm}
    \resizebox{0.8\columnwidth}{!}{
    \begin{tabular}{cccc}
    \toprule
    \textbf{Methods} &  \textbf{Modeling (s)} & \textbf{Solving (s)} & \textbf{Total (s)}\\
    \midrule
    {PPCC method} & 0.3893 & 0.6714 & 1.0607 \\
    {NPPCC method} & 0.3889 & 0.5572 & 0.9461 \\
    \bottomrule
\end{tabular}
}
\end{table}

\textcolor{black}{}
\begin{table}[t]
    \centering
    \footnotesize
    \caption{\textcolor{black}{computational time (IEEE-141 cases)}}
    \label{table: computational time (IEEE-141 cases)}
    %\vspace{-0.2cm}
    \resizebox{0.8\columnwidth}{!}{
    \begin{tabular}{cccc}
    \toprule
    \textbf{Methods} &  \textbf{Modeling (s)} & \textbf{Solving (s)} & \textbf{Total (s)}\\
    \midrule
    {PPCC method} & 0.9831 & 1.2103 & 2.1934 \\
    {NPPCC method} & 0.9498 & 0.7523 & 1.7021 \\
    \bottomrule
\end{tabular}
}
\end{table}

\vspace{-0.2cm}
\section{Conclusion and Prospects}
\label{Section V}
\subsection{\textcolor{black}{Conclusion}}
\textcolor{black}{This paper proposes a privacy-preserved optimal dispatch method to integrate the thermal flexibility of buildings into the power systems. First, we established a centralized optimal dispatch model incorporating BLAs. Then, the privacy-preserved algorithm that integrates the transformation
-based encryption, constraint relaxation, and constraint extension techniques is proposed. Theoretical proofs are provided to demonstrate the algorithm’s privacy guarantees. The proposed algorithm enables the DSO to obtain optimal scheduling results while protecting the private information of the BLA. Finally, numerical tests demonstrate the algorithm's performance on numerical precision, computational efficiency, and privacy-preservation capacity. The function of building thermal flexibility in reducing system operational costs is also validated.}

\subsection{\textcolor{black}{Prospects}}
\textcolor{black}{In the future, we plan to conduct further research in the following aspects.}

\textcolor{black}{
(1) \textit{The privacy-preserved allocation strategy for BLAs}:  When the BLA receives the dispatch signal from the DSO, it needs to formulate the optimal power allocation strategy based on the terminal load characteristics to maximize it profits. This process poses serious privacy issues to end-users. This issue warrants in-depth study.}

\textcolor{black}{
(2) \textit{Consideration on cybersecurity issues}: Cybersecurity issues like the risk of noisy data injection by hackers are dismissed in this paper. Future research could explore hybrid frameworks that combine privacy-preserved optimization with robust cybersecurity protocols to achieve comprehensive protection.}

\textcolor{black}{
\textit{(3) Privacy-preserved strategies considering uncertainty of occupants' behavior:} The inherent randomness in occupants' behavior can be propagated to the BLA, leading to uncertain parameters of the aggregate model. The privacy-preserved optimal dispatch considering occupants' uncertain activities deserve further research.
}
\appendices
\numberwithin{equation}{section}
\newcommand{\BIBdecl}{\setlength{\itemsep}{0.01 em}}
\bibliographystyle{IEEEtran}
\bibliography{IEEEabrv, References.bib}

\begin{thebibliography}{10}
\providecommand{\url}[1]{#1}
\csname url@samestyle\endcsname
\providecommand{\newblock}{\relax}
\providecommand{\bibinfo}[2]{#2}
\providecommand{\BIBentrySTDinterwordspacing}{\spaceskip=0pt\relax}
\providecommand{\BIBentryALTinterwordstretchfactor}{4}
\providecommand{\BIBentryALTinterwordspacing}{\spaceskip=\fontdimen2\font plus
\BIBentryALTinterwordstretchfactor\fontdimen3\font minus \fontdimen4\font\relax}
\providecommand{\BIBforeignlanguage}[2]{{%
\expandafter\ifx\csname l@#1\endcsname\relax
\typeout{** WARNING: IEEEtran.bst: No hyphenation pattern has been}%
\typeout{** loaded for the language `#1'. Using the pattern for}%
\typeout{** the default language instead.}%
\else
\language=\csname l@#1\endcsname
\fi
#2}}
\providecommand{\BIBdecl}{\relax}
\BIBdecl

\bibitem{wang2024robust}
G.~Wang, Y.~Zhou, Z.~Lin, S.~Zhu, R.~Qiu, Y.~Chen, and J.~Yan, ``Robust energy management through aggregation of flexible resources in multi-home micro energy hub,'' \emph{Applied Energy}, vol. 357, p. 122471, 2024.

\bibitem{zhong2023optimal}
J.~Zhong, Y.~Li, Y.~Wu, Y.~Cao, Z.~Li, Y.~Peng, X.~Qiao, Y.~Xu, Q.~Yu, X.~Yang \emph{et~al.}, ``Optimal operation of energy hub: An integrated model combined distributionally robust optimization method with stackelberg game,'' \emph{IEEE Transactions on Sustainable Energy}, vol.~14, no.~3, pp. 1835--1848, 2023.

\bibitem{taha2017buildings}
A.~F. Taha, N.~Gatsis, B.~Dong, A.~Pipri, and Z.~Li, ``Buildings-to-grid integration framework,'' \emph{IEEE Trans. Smart Grid}, vol.~10, no.~2, pp. 1237--1249, 2017.

\bibitem{hou2025robust}
Z.~Hou, S.~Lu, Z.~Wu, W.~Gu, H.~Zhang, Y.~Xu, and Z.~Gao, ``Robust parameter estimation of aggregate thermal dynamic model: A privacy-preserved approach,'' \emph{CSEE Journal of Power and Energy Systems}, 2025.

\bibitem{fontenot2019modeling}
H.~Fontenot and B.~Dong, ``Modeling and control of building-integrated microgrids for optimal energy management--a review,'' \emph{Appl. Energy}, vol. 254, p. 113689, 2019.

\bibitem{guo2024optimal}
Y.~Guo, Y.~Li, S.~Zhou, Z.~Zhang, Y.~Wang, Y.~Xu, X.~Yang, Z.~Li, and M.~Shahidehpour, ``Optimal dispatch for integrated energy system considering data-driven dynamic energy hubs and thermal dynamics of pipeline networks,'' \emph{IEEE Transactions on Smart Grid}, 2024.

\bibitem{niu2019flexible}
J.~Niu, Z.~Tian, Y.~Lu, and H.~Zhao, ``Flexible dispatch of a building energy system using building thermal storage and battery energy storage,'' \emph{Appl. Energy}, vol. 243, pp. 274--287, 2019.

\bibitem{song2018state}
M.~Song, C.~Gao, M.~Shahidehpour, Z.~Li, J.~Yang, and H.~Yan, ``State space modeling and control of aggregated tcls for regulation services in power grids,'' \emph{IEEE Trans. Smart Grid}, vol.~10, no.~4, pp. 4095--4106, 2018.

\bibitem{correa2019optimal}
C.~A. Correa-Florez, A.~Michiorri, and G.~Kariniotakis, ``Optimal participation of residential aggregators in energy and local flexibility markets,'' \emph{IEEE Trans. Smart Grid}, vol.~11, no.~2, pp. 1644--1656, 2019.

\bibitem{lu2021data}
S.~Lu, W.~Gu, S.~Ding, S.~Yao, H.~Lu, and X.~Yuan, ``Data-driven aggregate thermal dynamic model for buildings: A regression approach,'' \emph{IEEE Trans. Smart Grid}, vol.~13, no.~1, pp. 227--242, 2021.

\bibitem{zhang2019day}
R.~Zhang, T.~Jiang, W.~Li, G.~Li, H.~Chen, and X.~Li, ``Day-ahead scheduling of integrated electricity and district heating system with an aggregated model of buildings for wind power accommodation,'' \emph{IET Renewable Power Gener.}, vol.~13, no.~6, pp. 982--989, 2019.

\bibitem{tang2022multi}
H.~Tang and S.~Wang, ``Multi-level optimal dispatch strategy and profit-sharing mechanism for unlocking energy flexibilities of non-residential building clusters in electricity markets of multiple flexibility services,'' \emph{Renewable Energy}, vol. 201, pp. 35--45, 2022.

\bibitem{yang2021survey}
X.~Yang, L.~Shu, J.~Chen, M.~A. Ferrag, J.~Wu, E.~Nurellari, and K.~Huang, ``A survey on smart agriculture: Development modes, technologies, and security and privacy challenges,'' \emph{IEEE/CAA J. Autom. Sin.}, vol.~8, no.~2, pp. 273--302, 2021.

\bibitem{zhang2024admm}
P.~Zhang, S.~A. Mansouri, A.~R. Jordehi, M.~Tostado-V{\'e}liz, Y.~Z. Alharthi, and M.~Safaraliev, ``An admm-enabled robust optimization framework for self-healing scheduling of smart grids integrated with smart prosumers,'' \emph{Applied Energy}, vol. 363, p. 123067, 2024.

\bibitem{zhang2018admm}
C.~Zhang, M.~Ahmad, and Y.~Wang, ``Admm based privacy-preserving decentralized optimization,'' \emph{IEEE Trans. Inf. Forensics Secur.}, vol.~14, no.~3, pp. 565--580, 2018.

\bibitem{shang2021differentially}
F.~Shang, T.~Xu, Y.~Liu, H.~Liu, L.~Shen, and M.~Gong, ``Differentially private admm algorithms for machine learning,'' \emph{IEEE Trans. Inf. Forensics Secur.}, vol.~16, pp. 4733--4745, 2021.

\bibitem{liu2024differentially}
H.~Liu, S.~Lei, L.~Zhang, Y.~Huang, H.~Zhang, and C.~Peng, ``Differentially private distributed algorithm for energy sharing game with generalized demand bidding,'' in \emph{2024 IEEE Power \& Energy Society General Meeting (PESGM)}.\hskip 1em plus 0.5em minus 0.4em\relax IEEE, 2024, pp. 1--5.

\bibitem{huang2019dp}
Z.~Huang, R.~Hu, Y.~Guo, E.~Chan-Tin, and Y.~Gong, ``Dp-admm: Admm-based distributed learning with differential privacy,'' \emph{IEEE Trans. Inf. Forensics Secur.}, vol.~15, pp. 1002--1012, 2019.

\bibitem{dvorkin2020differentially}
V.~Dvorkin, F.~Fioretto, P.~Van~Hentenryck, P.~Pinson, and J.~Kazempour, ``Differentially private optimal power flow for distribution grids,'' \emph{IEEE Trans. Power Syst.}, vol.~36, no.~3, pp. 2186--2196, 2020.

\bibitem{wu2021privacy}
T.~Wu, C.~Zhao, and Y.-J.~A. Zhang, ``Privacy-preserving distributed optimal power flow with partially homomorphic encryption,'' \emph{IEEE Trans. Smart Grid}, vol.~12, no.~5, pp. 4506--4521, 2021.

\bibitem{shoukry2016privacy}
Y.~Shoukry, K.~Gatsis, A.~Alanwar, G.~J. Pappas, S.~A. Seshia, M.~Srivastava, and P.~Tabuada, ``Privacy-aware quadratic optimization using partially homomorphic encryption,'' in \emph{2016 IEEE 55th Conference on Decision and Control (CDC)}.\hskip 1em plus 0.5em minus 0.4em\relax IEEE, 2016, pp. 5053--5058.

\bibitem{xin2017information}
S.~Xin, Q.~Guo, J.~Wang, C.~Chen, H.~Sun, and B.~Zhang, ``Information masking theory for data protection in future cloud-based energy management,'' \emph{IEEE Trans. Smart Grid}, vol.~9, no.~6, pp. 5664--5676, 2017.

\bibitem{jia2022chance}
M.~Jia, G.~Hug, Y.~Su, and C.~Shen, ``Chance-constrained opf: A distributed method with confidentiality preservation,'' \emph{IEEE Trans. Power Syst.}, vol.~38, no.~4, pp. 3373--3387, 2022.

\bibitem{wu2017transformation}
L.~Wu, ``A transformation-based multi-area dynamic economic dispatch approach for preserving information privacy of individual areas,'' \emph{IEEE Trans. Smart Grid}, vol.~10, no.~1, pp. 722--731, 2017.

\bibitem{gonccalves2021critical}
C.~Gon{\c{c}}alves, R.~J. Bessa, and P.~Pinson, ``A critical overview of privacy-preserving approaches for collaborative forecasting,'' \emph{Int. J. Forecasting}, vol.~37, no.~1, pp. 322--342, 2021.

\bibitem{rigo2022iterative}
R.~Rigo-Mariani and V.~Vai, ``An iterative linear distflow for dynamic optimization in distributed generation planning studies,'' \emph{International Journal of Electrical Power \& Energy Systems}, vol. 138, p. 107936, 2022.

\bibitem{zhang2016robust}
C.~Zhang, Y.~Xu, Z.~Y. Dong, and J.~Ma, ``Robust operation of microgrids via two-stage coordinated energy storage and direct load control,'' \emph{IEEE Trans. Power Syst.}, vol.~32, no.~4, pp. 2858--2868, 2016.

\bibitem{Zeyin2025Parameters}
\BIBentryALTinterwordspacing
Z.~Hou and S.~Lu, ``Parameter settings: 33 and 141 bus distribution systems,'' 2025. [Online]. Available: \url{https://github.com/GreatTM/ZeyinHou--Data-for-Integrating-Building-Thermal-Flexibility-Into-Distribution-System}
\BIBentrySTDinterwordspacing

\end{thebibliography}

\begin{IEEEbiography}[{\includegraphics[width=1in,height=1.25in,clip,keepaspectratio]{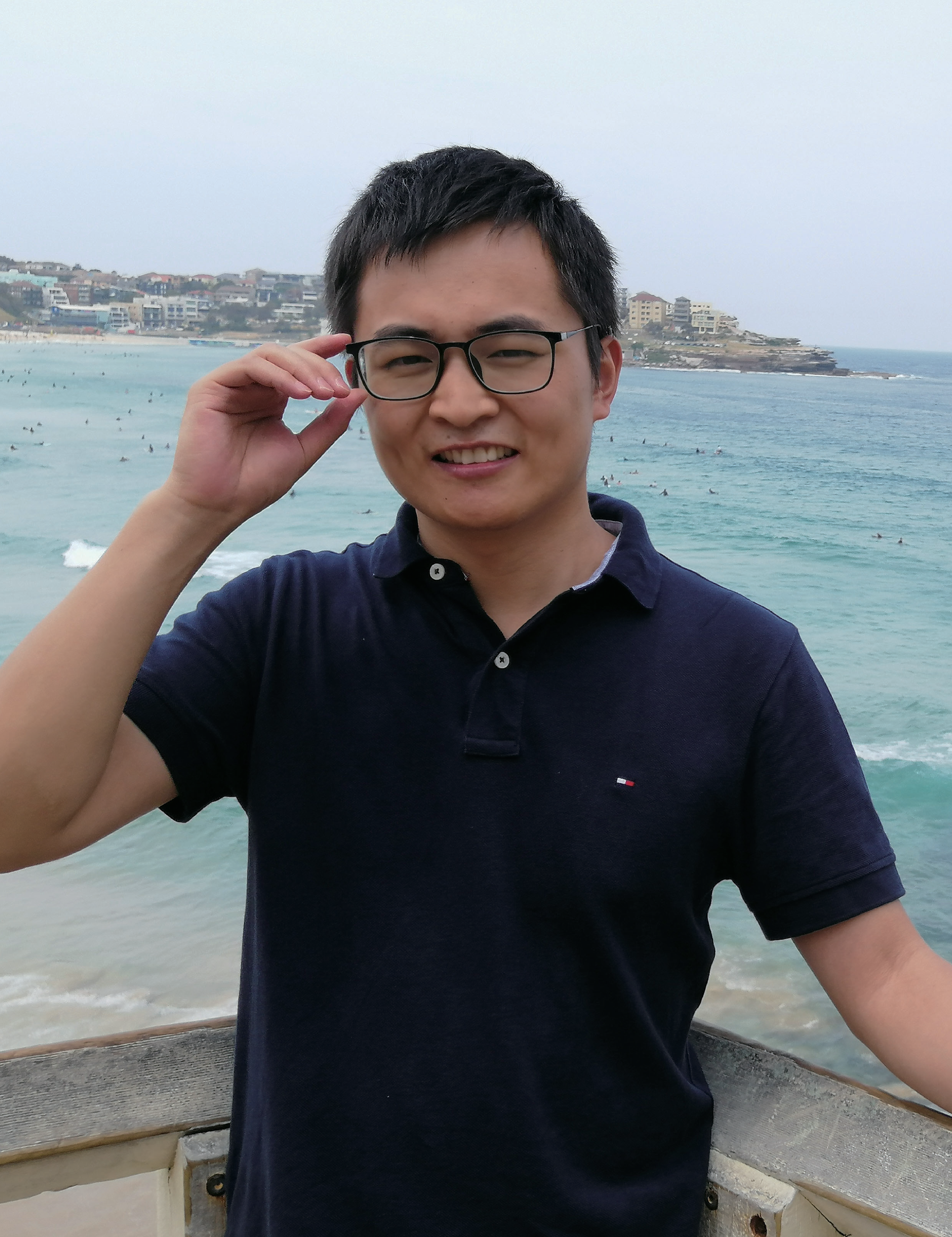}}]{Shuai Lu}(Member, IEEE)  is currently an Associate Professor at the School of Electrical Engineering, Southeast University, Nanjing China. He received his B.S. degree in Smart Grid Information Engineering from Nanjing University of Science and Technology, Nanjing, China in 2016 and his Ph.D. degree in Electrical Engineering from Southeast University, Nanjing China, in 2021. From 2018 to 2019, he was a visiting scholar at the University of New South Wales, Sydney, Australia.

He is a Young Editorial Board Member of \textit{Applied Energy} and \textit{Electric Power Automation Equipment}. His research interests include power and energy systems, operations research, and advanced computing technology.
\end{IEEEbiography}

\begin{IEEEbiography}[{\includegraphics[width=1in,height=1.25in,clip,keepaspectratio]{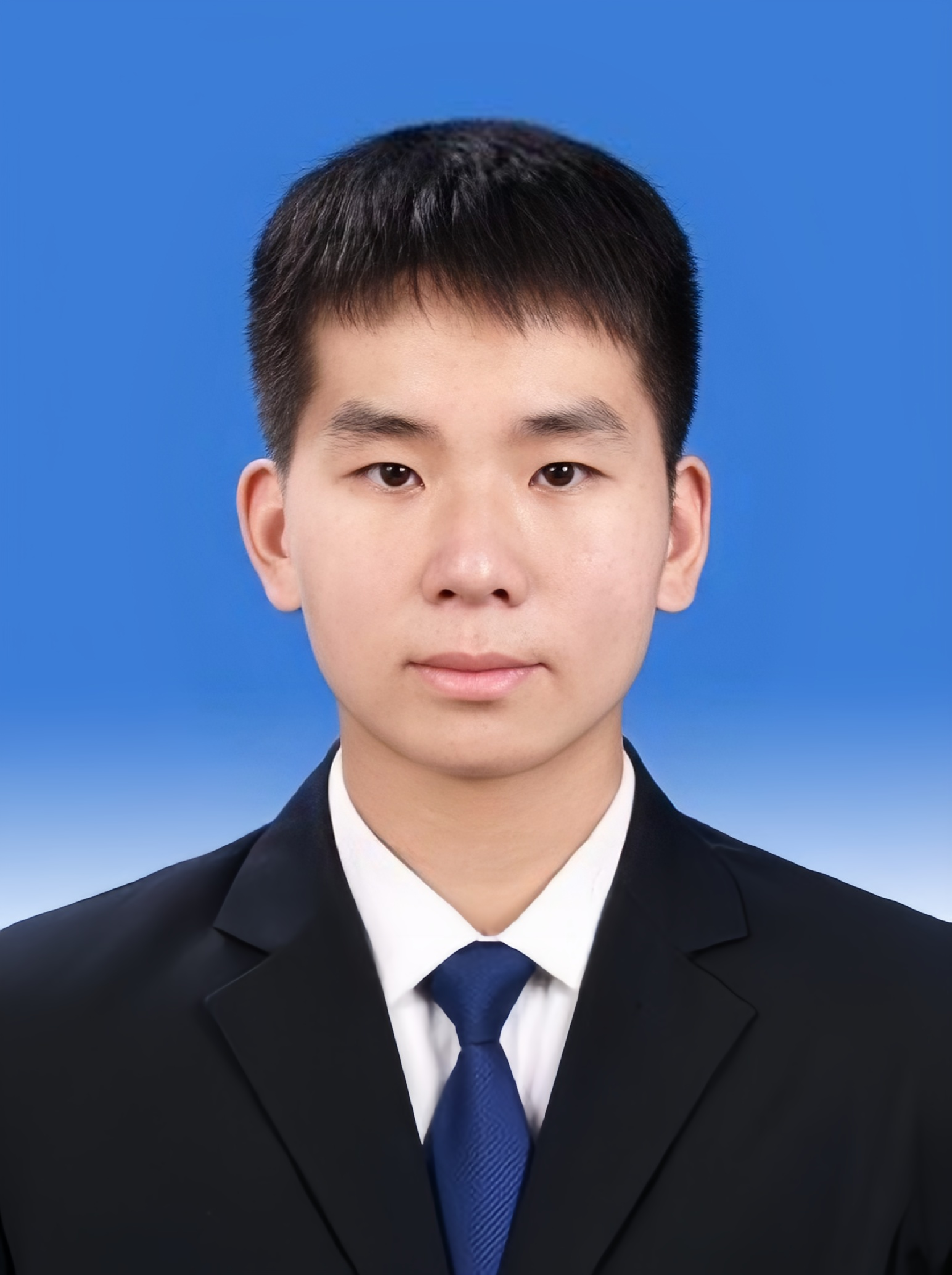}}]{Zeyin Hou}(Student Member, IEEE)  received the B.S. degree in Electrical Engineering from Xi'an Jiaotong University, Xi'an, China, in 2022. He is currently working toward the M.S. degree in the School of Electrical Engineering from Southeast University, Nanjing, China.

His research interests include privacy-preserving and data-driven techniques in power systems.
\end{IEEEbiography}

\begin{IEEEbiography}[{\includegraphics[width=1in,height=1.25in,clip,keepaspectratio]{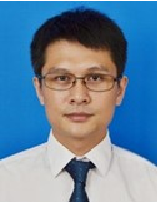}}]{Wei Gu}(Senior Member, IEEE)  received his B.S. and Ph.D. degrees in Electrical Engineering from Southeast University, China, in 2001 and 2006, respectively. From 2009 to 2010, he was a Visiting Scholar in the Department of Electrical Engineering, Arizona State University. 

He is now a professor at the School of Electrical Engineering, Southeast University. He is the director of the institute of distributed generations and active distribution networks. His research interests include distributed generations and microgrids, integrated energy systems. He was an Editor for the IEEE Transactions on Power Systems, the IET Energy Systems Integration and the Automation of Electric Power Systems (China). 
\end{IEEEbiography}

\begin{IEEEbiography}[{\includegraphics[width=1in,height=1.25in,clip,keepaspectratio]{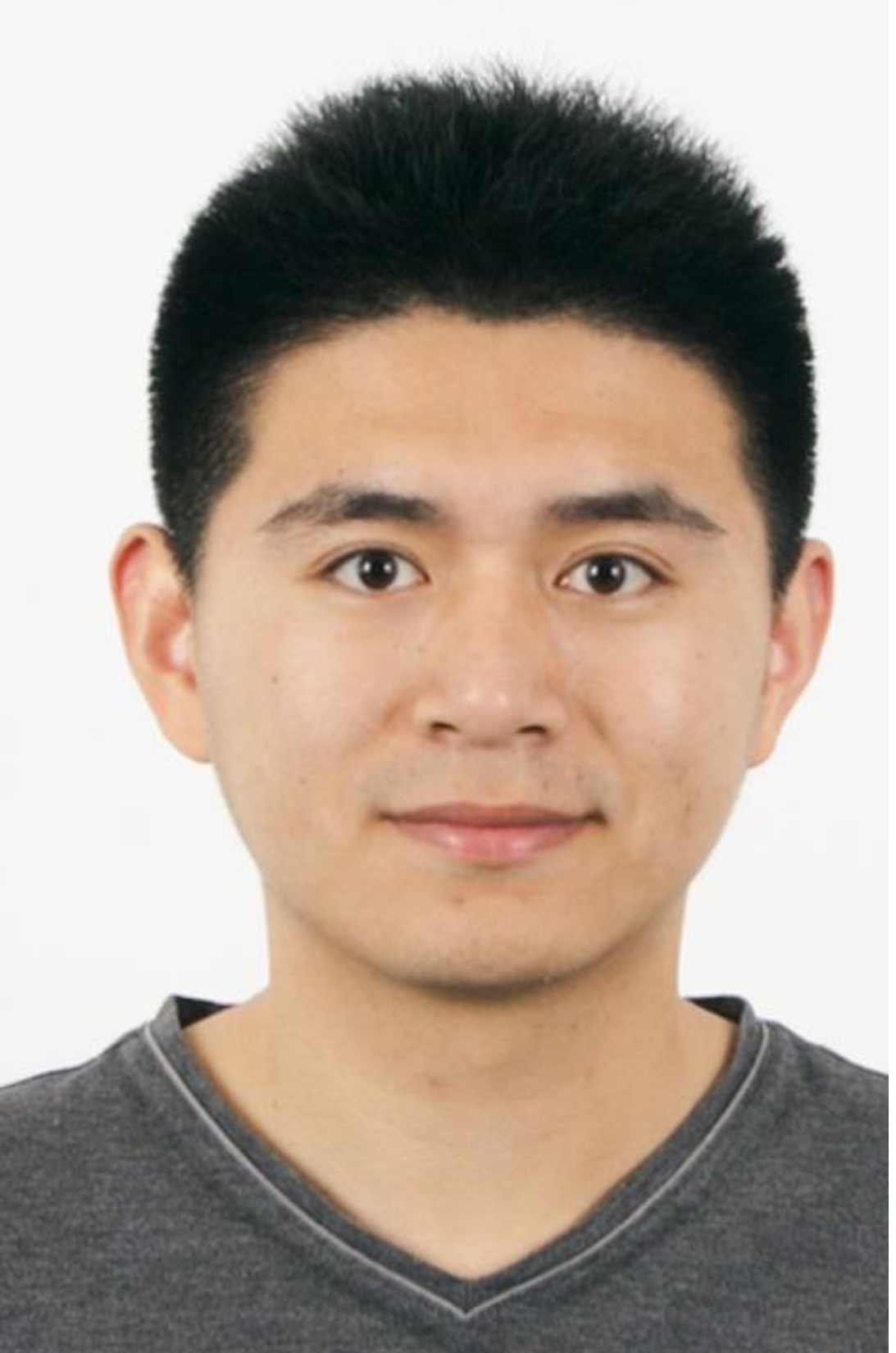}}]{Yijun Xu}(Senior Member, IEEE) is a professor at Southeast University, Nanjing, China. He received his Ph.D. degree from the Bradley Department of Electrical and Computer Engineering at Virginia Tech, Falls Church, VA, in December 2018. He worked as a research assistant professor at Virginia Tech-Northern Virginia Center, Falls Church, VA, in 2021. He was a postdoc associate at the same institute from 2019 to 2020.  
He did a computation internship at Lawrence Livermore  National Laboratory, Livermore, CA, and a power engineer internship at ETAP -- Operation Technology, Inc., Irvine, California, in 2018 and 2015, respectively. 

His research interests include power system uncertainty quantification, uncertainty inversion, and decision-making under uncertainty. 
Dr. Xu is currently serving as an Associate Editor of the \textsc{IET Generation, Transmission \& distribution}, an Associate Editor of the \textsc{IET Renewable Power Generation}, and the Young Editor of the \textsc{Power System Protection and Control}. He is the co-chair of the IEEE Task Force on Power System Uncertainty Quantification and Uncertainty-Aware Decision-Making.
\end{IEEEbiography}

\end{document}